\documentclass[preprint,showpacs,preprintnumbers,amsmath,amssymb]{revtex4}

\usepackage{graphicx,bm}
\usepackage{here}

\begin{document}

\preprint{HUPD-0704}

\title{
$\pi$ and $\sigma$ mesons at finite temperature and density in the NJL
model with dimensional regularization}

\author{T. Inagaki}
 \affiliation{Information Media Center, Hiroshima University, 
   Higashi-Hiroshima, Japan}
\author{D. Kimura} 
 \affiliation{Department of Physical Science, Hiroshima University,
   Higashi-Hiroshima, Japan}
\author{A. Kvinikhidze}
 \affiliation{A. Razmadze Mathematical Institute of Georgian Academy of 
   Sciences, Tbilisi, Georgia}

\date{\today}

\begin{abstract}
Dynamical Symmetry breaking and 
meson masses are studied in
the Nambu-Jona-Lasinio (NJL) model at finite temperature and 
chemical potential 
using the dimensional regularization.
Since the model is not renormalizable in four space-time 
dimensions, physical results and parameters
depend on the regularization 
method. Following the imaginary time formalism, we introduce
the temperature, $T$ and the chemical potential, $\mu$.
The parameters of the model are fixed by calculating 
the pion mass and decay constant in the dimensional regularization
at $T=\mu=0$.
\end{abstract}


\maketitle

\section{Introduction}
QCD is a fundamental theory of quarks and gluons
whose 
coupling constant 
is large at low energy scale.
Therefore one cannot adopt the perturbative expansion in 
powers of the 
coupling constant at low energy. 
To study physics in a hadronic phase we can not avoid considering
the non-perturbative effect of QCD. One of the possible ways to evaluate
the phenomena in the hadronic phase is to use a low energy effective
theory. Some phenomenological parameters are introduced to construct the 
low energy effective theory which is simpler than QCD to deal with. NJL model is one of the low energy effective 
theories of QCD \cite{NJL}, for a review see Refs.\cite{H,RWP,V,HK}.
The model has the same chiral symmetry as QCD and the symmetry is broken 
down dynamically. Thus it is often used to study 
symmetry properties of 
QCD in the hadronic phase.

The NJL model contains four-fermion interactions. Since a four-fermion 
interaction is a dimension six operator, the NJL model is not renormalizable
in four space-time dimensions. 
To obtain finite expressions 
we must regularize the theory. In such a non-renormalizable model
most of the physical quantities depend on the regularization method and a 
parameter introduced to regularize the theory. For example, a cut-off scale 
is introduced as a parameter in a cut-off regularization. The coupling 
constant and the cut-off scale are determined phenomenologically.

The regularization 
in general may break some of the 
symmetries 
of the theory.
For example a naive  
three dimensional cut-off breaks 
Lorentz and gauge 
invariance
 \cite{AIBY}. In the present paper we employ the 
dimensional regularization. 
It preserves most of the 
 symmetries of the theory, including the general 
covariance.
We regard the space-time dimensions as one of 
the parameters in the 
effective theory. Thus the dimensions should be determined in some
low energy phenomena which have some relation with chiral symmetry 
breaking.

There are some works
investigating 
 NJL model in the dimensional 
regularization. 
The general properties of the renormalization and the renormalization
group is studied in arbitrary dimensions $2<D<4$ \cite{KY, M, HKWY}.
In  Refs.\cite{IKM, IMM, II, I, IMO, II2, SC} 
the NJL model is considered as a prototype model of composite Higgs.
The phase structure of dynamical symmetry breaking is analyzed at 
high temperature, density, electro-magnetic field and curvature. 
But we have no established fundamental theory of composite Higgs models 
at high energy scale. Thus the physical scale of the theory has not been
fixed and the contribution of the current fermion mass has not been 
analyzed. 

In Ref.\cite{KN} the dimensional regularization is 
modified to keep
 four dimensional properties of the non-renormalizable theory as much as possible.
To achieve this goal the dimensional regularization is applied to only the radial
part in loop integrals. It is one of the analytic regularization.  
The meson loop contribution to the chiral symmetry breaking is also analyzed 
in the NJL model 
with the modified dimensional regularization \cite{JR}.
The procedure is easily extended to the finite temperature and chemical
potential. However, the phenomenologically consistent dimensions are 
less than two in this approach. Because of the infrared divergence it 
seems to be difficult to obtain a finite result at finite temperature.
 
In the present paper we regard the NJL model as a low energy 
effective theory of QCD and apply the dimensional regularization not only
to 
the radial part but also 
to 
the angular parts of internal 
momenta
in loop integrals. We include the 
effect of the current quark mass and 
fix the scale through the observed properties of pseudo-scalar mesons, 
the pion mass and the decay constant 
at zero density and temperature.
The constituent quark mass and the meson properties are evaluated in 
thermal equilibrium.

The paper is organized in the following way. First we apply the dimensional 
regularization to the NJL model with two flavors of quarks.
As is well-known, the small mass of quarks explicitly breaks the chiral 
symmetry. Thus the chiral symmetry is only 
approximate.
In such a model we evaluate the phase structure of the theory. 
We need to perform renormalization as well to define a positive coupling 
constant. 
In Sec.II we calculate the mass of scalar and pseudo-scalar mesons. We take
the massless quarks limit and show that the dimensional regularization keeps 
the Nambu-Goldstone modes massless. A relationship between the 
space-time dimensions and the physical mass scale 
is also discussed. 
In Sec.IV we introduce the thermal effect 
in the imaginary time
formalism and evaluate the effective potential. 
In Sec.V we calculate the mass of scalar and pseudo-scalar mesons at 
finite $T$ and $\mu$.
The three massless poles of pseudo-scalar 
mesons survive at higher temperature and chemical potential. 
At the end concluding remarks are given.

\section{Dynamical symmetry breaking}
The chiral symmetry is broken when the composite operator
of a quark and an anti-quark, $\bar{\psi}\psi$, develops a 
non-vanishing expectation value. It is caused by the QCD
dynamics and 
is 
called dynamical symmetry breaking.

The NJL model is one of the simplest 
models
of dynamical chiral symmetry breaking.
The model is originally introduced to evaluate the low
energy properties of hadrons as a bound state of some
primary fermion fields \cite{NJL}.
The Lagrangian of the two-flavor NJL model is defined by
\begin{equation}
 {\cal L} = \bar{\psi} (i 
       \partial\hspace{-2.9mm}{\not}\hspace{2.9mm} -m) \psi
       +g_\pi^0 \left\{ (\bar\psi \psi)^2 
         + ( \bar\psi i\gamma_5\tau^a \psi)^2 \right\} \ ,
\label{NJL-lag}
\end{equation}
where $g_\pi^0$ is an effective coupling constant, $\tau^a$ represents
the isospin Pauli matrices, and $m=\mbox{diag}(m_u, m_d)$ 
is the mass matrix of up and down quarks. 
In the Lagrangian (\ref{NJL-lag}) we omit the color and the flavor indeces.

In the limit of massless quarks this Lagrangian is invariant 
under the global flavor transformation
\begin{equation}
  \psi \rightarrow e^{i\theta^a \tau^a}\psi ,
\end{equation}
and the chiral transformation
\begin{equation}
  \psi \rightarrow e^{i\theta^a \tau^a \gamma^5}\psi .
\end{equation}
The quark mass term violates these symmetries explicitly.
When the composite operator $\bar{\psi}\psi$ develops a non-vanishing
expectation value, the quark acquires a mass and the chiral symmetry is 
broken dynamically.

The vacuum expectation value of the composite operator 
$\bar{\psi}\psi$ can be found by solving the gap equation.
The gap equation of the NJL model in $D$ dimensions is

\begin{equation}
  \langle\sigma\rangle  =  2 i g_\pi^0 \int \frac{d^D k}{(2\pi)^D} \mbox{tr} S(k),
\label{eq:gap}
\end{equation} 
where $S(k)$ is the fermion propagator and "tr" denotes trace with respect to flavor, 
color and spinor indices,
\begin{equation}
  S(k) \equiv \frac{1}{k\hspace{-2.9mm}{\not}\hspace{2.9mm}-m-\langle\sigma\rangle+i\epsilon}.
\end{equation}
Integration over $k$ takes the gap equation to the form
\begin{eqnarray}
    \langle\sigma\rangle &=& \frac{2 N_c g_\pi^0}{(2\pi)^{D/2}}
          \Gamma\left(1-\frac{D}{2}\right)\sum_{j\in \{u,d\}}
          (m_j+\langle\sigma\rangle)[(m_j+\langle\sigma\rangle)^2]^{(D/2-1)}
\nonumber \\
    & \equiv &  g_\pi^0 A(D) \sum_{j\in \{u,d\}} (m_j+\langle\sigma\rangle) [(m_j+\langle\sigma\rangle)^2]^{(D/2-1)},
\label{gap}
\end{eqnarray}
where $N_c$ is the number of colors
and $A(D)$ is defined by
\begin{equation}
  A(D) \equiv \frac{2 N_c}{(2\pi)^{D/2}}
           \Gamma\left(1-\frac{D}{2}\right) .
\end{equation}
The gap equation (\ref{gap}) has three solutions. 
To find the stable solution we evaluate the effective potential for the
scalar channel, $\sigma$, (see, for example Ref.\cite{IKM})
\begin{eqnarray}
    V(\sigma)= \frac{\sigma^2}{4 g_\pi^0} -  
    \frac{A(D)}{2D} \sum_{j\in \{u,d\}} [(m_j+\sigma)^2]^{D/2}.
\label{eff}
\end{eqnarray}
Extrema of the effective potential satisfy the gap equation
(\ref{gap}).
The most stable solution is determined by the minimum
of the effective potential. 
\begin{figure}[t]
\begin{center}
\includegraphics{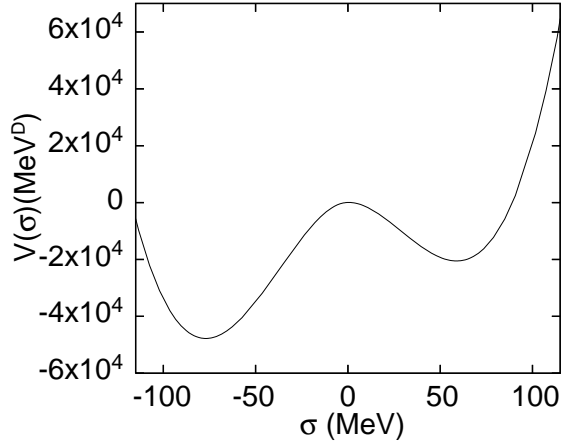}
\caption{Typical behavior of the effective potential ($D=2.8$, $m_u=3$MeV,
$m_d=5$MeV and $g_\pi^0=-0.01$MeV$^{2-D}$).}
\label{epotd28}
\end{center}
\end{figure}
As is shown in the 
Fig.\ref{epotd28}, we find the minimum at a negative value of 
$\sigma$ for a negative coupling constant, $g_\pi^0$, with a finite current 
quark mass. We note that the expectation value,
$\displaystyle \langle\bar{\psi}\psi\rangle \sim -\frac{1}{2 g_\pi^0} \langle\sigma\rangle$, has a negative value as usual.

At the massless limit, $m\rightarrow 0$, the gap equation (\ref{gap}) is
simplified and 
the solutions are:
\begin{equation}
  \langle\sigma\rangle=0,
\label{sol1:gap}
\end{equation}
\begin{equation}
  (\langle\sigma\rangle^2)^{D/2-1}=\frac{1}{2 g_\pi^0 A(D)}.
\label{sol2:gap}
\end{equation}
The non-vanishing solution (\ref{sol2:gap}) is stabler, see Fig.1. 
It shows that the composite operator $\bar{\psi}\psi$ develops a 
non-vanishing expectation value and the chiral symmetry is broken 
dynamically. It should be noted that
\begin{equation}
  A(D) < 0,\ \ \mbox{for}\ \ 2 < D < 4 .
\end{equation}
The left hand side of 
the 
Eq.(\ref{sol2:gap}) is real and positive.
To find a real solution of Eq.(\ref{sol2:gap}) we must set a negative 
value for the coupling constant $g_\pi^0$. It is one of characteristic 
features of the dimensional regularization.

\subsection{Renormalization}
The Lagrangian (1) is not renormalizable 
for $D=4$. 
However, the renormalization is 
sometimes useful to connect
the results in the different regularization schemes.
Here we renormalize parameters in our model.

In the leading order of the $1/N$ expansion the radiative correction
of the four-fermion coupling for 
the scalar channel is given by the
summation of all bubble type diagrams,
\begin{eqnarray}\label{112}
  G_s (p^2,\langle\sigma\rangle) &\equiv&
\begin{minipage}{20mm}
\hspace*{1mm}
  \includegraphics[height=16.8mm]{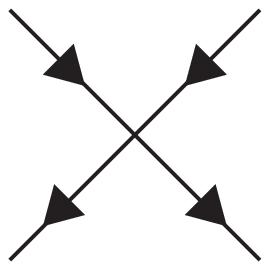}
\end{minipage}
  +
\begin{minipage}{40mm}
\hspace*{1mm}
  \includegraphics[height=16.8mm]{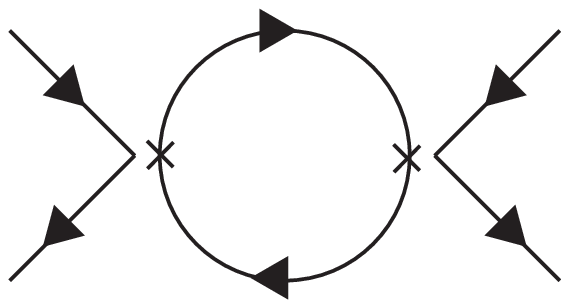}
\end{minipage}
\nonumber \\
  && +
\begin{minipage}{55mm}
\hspace*{1mm}
  \includegraphics[height=16.8mm]{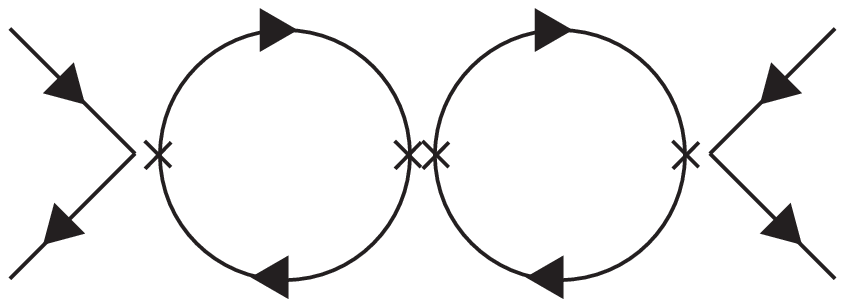}
\end{minipage}
\nonumber \\
  && +
\begin{minipage}{68mm}
\hspace*{1mm}
  \includegraphics[height=16.8mm]{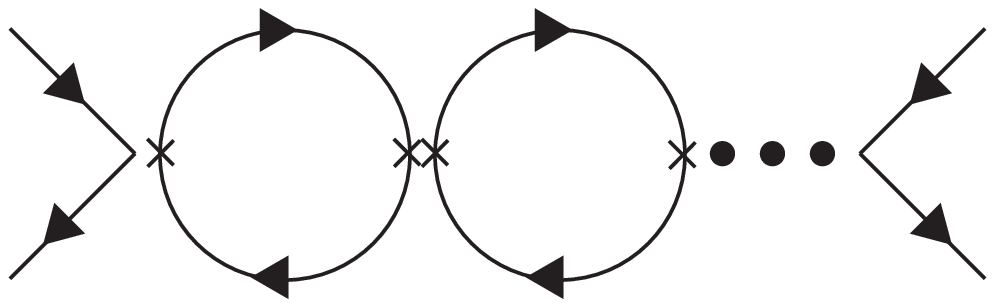}
\end{minipage}
\nonumber \\
  & = & 2 g_\pi^0 \left( 1 + \frac{\Pi_{s}(p^2)}
       {2 g_\pi^0 -\Pi_{s}(p^2)}\right)
  = \frac{4 (g_\pi^0)^2}
       {2 g_\pi^0 -\Pi_{s}(p^2)},
\end{eqnarray}
where the self-energy $\Pi_{s}(p^2)$ is
\begin{eqnarray}
  \Pi_{s}(p^2) &=&
\begin{minipage}{20mm}
\hspace*{1mm}
  \includegraphics[height=16.8mm]{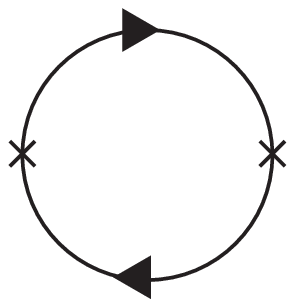}
\end{minipage}
  =  4 i (g_\pi^0)^2 \int \frac{d^D k}{(2\pi)^D} 
  \mbox{tr} [S(k)S(k-p)]
\nonumber \\
  &=& 2(g_\pi^0)^2 (D-1)A(D)\sum_{j\in \{u,d\}} \int^1_0 dx L_j^{D/2-1},
\label{pis}
\end{eqnarray}
with $L_j=(m_j+\langle\sigma\rangle)^2-x(1-x)p^2$.

We renormalize the coupling constant $g_\pi^0$ by imposing the
following renormalization condition
\begin{eqnarray}
  G_s (p^2=0,\langle\sigma\rangle=M_0) & \equiv & Z_{g}(M_0) G_s^r (p^2=0,\langle\sigma\rangle=M_0)
\nonumber \\
  & = & \frac{4 Z_{g}(M_0) g_\pi^0}{\displaystyle \sum_{j\in \{u,d\}}
  [(m_j+M_0)^2]^{D/2-1}}
\nonumber \\ 
  & = & \frac{4 g_\pi^r}{\displaystyle \sum_{j\in \{u,d\}}
  [(m_j+M_0)^2]^{D/2-1}} ,
\label{cond:ren}
\end{eqnarray}
where we introduce the renormalization constant, $Z_{g}(M_0)$. The superscript
$r$ stands for the renormalized quantities and $M_0$ is a renormalization 
scale. 

At the limit $p^2=0$ the self-energy $\Pi_{s}(p^2)$ simplifies to
\begin{equation}
  \Pi_{s}(p^2=0)|_{\langle\sigma\rangle=M_0} = 
  2 (g_\pi^0)^2 (D-1) A(D) \sum_{j\in \{u,d\}} [(m_j+M_0)^2]^{D/2-1}.
\label{pisp0}
\end{equation}
Thus the renormalization constant is given by
\begin{equation}
  Z_g(M_0)=\frac{1}{2}\frac{\sum_{j\in \{u,d\}}[(m_j+M_0)^2]^{D/2-1}}
  {1-g^0_\pi (D-1) A(D) \sum_{j\in \{u,d\}}[(m_j+M_0)^2]^{D/2-1}} .
\end{equation}
Therefore the renormalized coupling $g_\pi^r$ is found to be
\begin{equation}
   \frac{1}{g_\pi^0}\frac{1}{\displaystyle \sum_{j\in \{u,d\}} [(m_j+M_0)^2]^{D/2-1}}
   = \frac{1}{2 g_\pi^r} + (D-1) A(D).
\label{def:reng}
\end{equation}

We rewrite the gap equation (\ref{gap}) in terms of the renormalized 
coupling constant $g_\pi^r$ as
\begin{equation}
  \frac{\displaystyle \sum_{j\in \{u,d\}} (m_j+\langle\sigma\rangle)[(m_j+\langle\sigma\rangle)^2]^{D/2-1}}
       {\displaystyle \sum_{j\in \{u,d\}} [(m_j+M_0)^2]^{D/2-1}}
  =\langle\sigma\rangle\left(\frac{1}{2g_\pi^r A(D)} + D-1 \right).
\label{gap2}
\end{equation}
In the massless limit 
$m\rightarrow 0$, the non-trivial solution of the
gap equation is given by
\begin{equation}
  (\langle\sigma\rangle^2)^{(D/2-1)}=\left(\frac{1}{2 g_\pi^r A(D)} 
                  + D-1 \right)(M_0^2)^{(D/2-1)}.
\end{equation}
A real solution for $\langle\sigma\rangle$ exists for 
$1/g_\pi^{r} < 1/g_\pi^{cr}$,
\begin{equation}
  g_\pi^{cr} = \frac{1}{2 (1-D) A(D)} > 0 \ \ \mbox{for}\ \ 2 < D < 4 .
\end{equation}
The critical coupling is real and positive. It should be noted that
the broken phase is realized for $g_{\pi}^r > g_\pi^{cr} > 0$ or
$g_\pi^r < 0$.

It may be useful to rewrite the Eqs. (\ref{def:reng}) and (\ref{gap2})
in terms of the critical coupling,
\begin{equation}
   \frac{1}{g_\pi^0}\frac{1}{\displaystyle \sum_{j\in \{u,d\}} [(m_j+M_0)^2]^{D/2-1}}
   = \frac{1}{2g_\pi^r} -\frac{1}{2g_\pi^{cr}},
\label{def:reng2}
\end{equation}
and
\begin{equation}
  \frac{\displaystyle \sum_{j\in \{u,d\}} (m_j+\langle\sigma\rangle)[(m_j+\langle\sigma\rangle)^2]^{1-D/2}}
       {\displaystyle \sum_{j\in \{u,d\}} [(m_j+M_0)^2]^{1-D/2}}
  =\frac{2\langle\sigma\rangle}{A(D)}\left(\frac{1}{g_\pi^r} - \frac{1}{g_\pi^{cr}} \right).
\label{gap3}
\end{equation}
From Eq.(\ref{def:reng2}) it is 
clear that the bare coupling 
$g_\pi^0$ should be negative in the broken phase, $1/g_\pi^r < 1/g_\pi^{cr}$,
in the dimensional regularization.

The renormalization group $\beta$ function is defined 
by \cite{M,HKWY},
\begin{equation}
  \beta(g_\pi^r) \equiv M_0 \left.
                            \frac{\partial g_\pi^r}{\partial M_0}
                            \right|_{g_\pi^0} .
\label{beta}
\end{equation}
Using Eq.(\ref{gap3}) 
in Eq.(\ref{beta}) we obtain 
\begin{equation}
  \beta(g_\pi^r) 
  =2(D-2)
  \frac{\displaystyle M_0\sum_{j\in \{u,d\}}[(m_j+M_0)^2]^{(D-3)/2}}
       {\displaystyle \sum_{j\in \{u,d\}}[(m_j+M_0)^2]^{D/2-1}}
  \frac{g_\pi^r (g_\pi^{cr}-g_\pi^r)}{g_\pi^{cr}}.
\end{equation}
The ultraviolet stable fixed point appears at the critical
coupling, $g_\pi^{r}=g_\pi^{cr}$.

\section{Meson masses}

To describe a bound state in 
the hadronic phase  non-perturbative effects 
of QCD should be considered. 
Here we study meson masses in the NJL model at the leading order 
of the $1/N$ expansion.

\subsection{Pseudo-scalar mesons}
First we calculate the pion mass in the two-flavor NJL model.
It corresponds to the Nambu-Goldstone modes at the massless quark 
limit. Since the current quark mass explicitly breaks the chiral 
symmetry, the pion mass should be proportional to the current 
quark mass in the broken phase.

The pion mass is defined by the pole of the propagator for
a pseudo-scalar channel. The leading order of the $1/N$ expansion
is described by the summation of infinite number of the bubble 
diagrams, 
\begin{eqnarray}
  G_{5}^{ab}(p^2,\langle\sigma\rangle) &=&
\begin{minipage}{30mm}
\vglue -2mm
  \includegraphics[height=6.6mm]{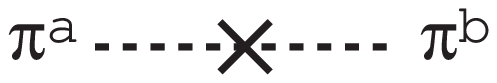}
\end{minipage}
  +
\begin{minipage}{50mm}
\hspace*{1mm}
  \includegraphics[height=16.8mm]{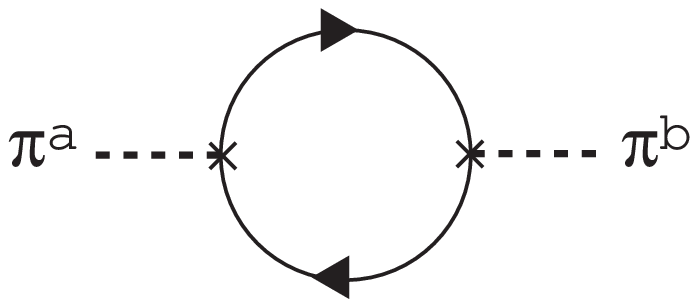}
\end{minipage}
\nonumber \\
  && + 
\begin{minipage}{56mm}
\hspace*{1mm}
  \includegraphics[height=16.8mm]{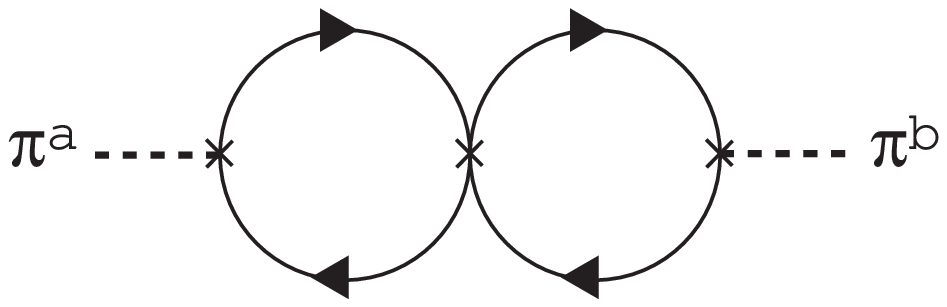}
\end{minipage}
\nonumber \\
  && + 
\begin{minipage}{68mm}
\hspace*{1mm}
  \includegraphics[height=16.8mm]{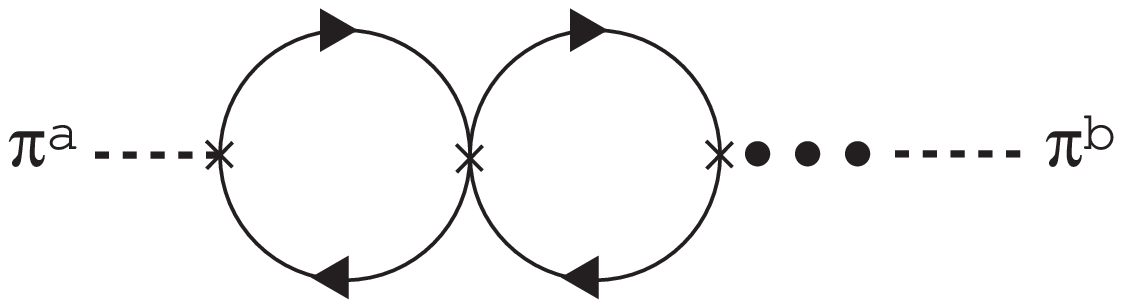}
\end{minipage}
\nonumber \\
  & = & \frac{4 (g_\pi^0)^2}{2 g_\pi^0 - \Pi_{5}^{a}(p^2)}\delta^{ab}
  \equiv \frac{Z_\pi^a M_0^{4-D}}{m^2_{\pi^a}-p^2+O(p^4)}\delta^{ab},
\label{pro:pi}
\end{eqnarray}
where we insert the renormalization scale $M_0^{D/2-2}$ to impose the 
correct mass dimensions on $Z_\pi$. 
$\Pi_{5}^{a}(p^2)$ is the self-energy for the pseudo-scalar 
channel which is given by
\begin{eqnarray}
  \Pi_{5}^{a}(p^2) & = & i\gamma^5 \tau^a
\begin{minipage}{20mm}
\hspace*{0.5mm}
  \includegraphics[height=16.8mm]{pis.eps}
\end{minipage}
  i\gamma^5 \tau^a
\nonumber \\
  & = & 4 i (g_\pi^0)^2 \int \frac{d^D k}{(2\pi)^D} 
  \mbox{tr} [i\gamma^5 \tau^a S(k) i\gamma^5 \tau^a S(k-p)] .
\label{pi5}
\end{eqnarray}
Integrating over the $D$-dimensional momentum $k$, we obtain
\begin{eqnarray}
  \Pi_{5}^{1,2}(p^2) &=&
  4 (g_\pi^0)^2 A(D) \int^1_0 dx \left[
    (1-D)x(1-x)p^2 + \frac{D}{2}x(m_u+\langle\sigma\rangle)^2\right.
\nonumber \\
  && \left. + \frac{D}{2}(1-x)(m_d+\langle\sigma\rangle)^2
    +\left(1-\frac{D}{2}\right)(m_u+\langle\sigma\rangle)(m_d+\langle\sigma\rangle)
  \right]
\nonumber \\
  && \times
   \left[x(m_u+\langle\sigma\rangle)^2+(1-x)(m_d+\langle\sigma\rangle)^2-x(1-x)p^2\right]^{D/2-2}, 
\nonumber \\
\end{eqnarray}
and
\begin{eqnarray}
  \Pi_{5}^{3}(p^2) &=& 2 (g_\pi^0)^2 A(D) \int^1_0 dx 
    \sum_{j\in \{u,d\}}\left[
    (1-D)x(1-x)p^2 +(m_j+\langle\sigma\rangle)^2
  \right]
\nonumber \\
  && \times \left[(m_j+\langle\sigma\rangle)^2-x(1-x)p^2\right]^{D/2-2}.
\end{eqnarray}

We can analytically calculate the momentum integral at finite $p^2$, 
but it is enough to evaluate the self-energy near $p^2=0$ to observe 
the massless pole of the Nambu-Goldstone mode. 
Straightforward calculations lead to
\begin{eqnarray}
  \Pi_{5}^{1,2}(p^2\sim 0) &=& 4 (g_\pi^0)^2 A(D)\left[ 
  \frac{\sum_{j\in \{u,d\}}
  [(m_j+\langle\sigma\rangle)^2]^{(D-1)/2}}
  {\sum_{j\in \{u,d\}}
  [(m_j+\langle\sigma\rangle)^2]^{1/2}}
  -f(D)p^2 \right]
\nonumber\\
  &&+O(p^4),
\end{eqnarray}
and
\begin{eqnarray}
  \Pi_{5}^{3}(p^2\sim 0) &=& (g_\pi^0)^2 A(D) \left[ 2 \sum_{j\in \{u,d\}}
  [(m_j+\langle\sigma\rangle)^2]^{D/2-1} \right.
\label{pi3}
\\ \nonumber
 && \left.
  +\left(1-\frac{D}{2}\right)p^2\sum_{j\in \{u,d\}}
  [(m_j+\langle\sigma\rangle)^2]^{D/2-2} \right]
  +O(p^4),
\end{eqnarray}
where $f(D)$ is given by
\begin{eqnarray}
  &&\hspace*{-10mm}f(D)
\nonumber \\ 
&& =\frac{1}{\left[(m_d+\langle\sigma\rangle)^2-(m_u+\langle\sigma\rangle)^2\right]^3}
\nonumber \\ && \times
\{
\left(\frac{2}{D}-1\right)\left[(m_d+\langle\sigma\rangle)^{D+2}-(m_u+\langle\sigma\rangle)^{D+2}\right]
\nonumber \\ &&
+\left(\frac{2}{D}+1\right)
\left[(m_d+\langle\sigma\rangle)^{D}(m_u+\langle\sigma\rangle)^{2}-(m_d+\langle\sigma\rangle)^{2}(m_u+\langle\sigma\rangle)^{D}\right]
\nonumber \\ &&
-\left(\frac{4}{D}-1\right)
\left[(m_d+\langle\sigma\rangle)^{D+1}(m_u+\langle\sigma\rangle)-(m_d+\langle\sigma\rangle)(m_u+\langle\sigma\rangle)^{D+1}\right]
\nonumber \\ && 
-(m_d+\langle\sigma\rangle)^{D-1}(m_u+\langle\sigma\rangle)^{3}+(m_d+\langle\sigma\rangle)^{D-1}(m_u+\langle\sigma\rangle)^{3}
\} .
\end{eqnarray}
In the isospin symmetric case, $m_u=m_d$,
$\Pi_{5}^{1,2}(p^2\sim 0)$ coincides with $\Pi_{5}^{3}(p^2\sim 0)$.

Further we obtain the  pion wave function
renormalization constants,
\begin{equation}
  Z_{\pi^{1,2}}^{-1} = -A(D)f(D)M_0^{4-D} ,
\end{equation}
and
\begin{equation}
  Z_{\pi^3}^{-1} = \frac{A(D)}{4}\left(1-\frac{D}{2}\right)M_0^{4-D}
  \sum_{j\in \{u,d\}}[(m_j+\langle\sigma\rangle)^2]^{D/2-2}.
\end{equation}
Neglecting O$(p^4)$ terms, the pion masses are found to be
\begin{equation}
   Z_{\pi^{1,2}}^{-1}M_0^{D-4}
   m_{\pi^{1,2}}^2 = \frac{1}{2 g_\pi^0} - A(D) 
  \frac{(\langle\sigma\rangle+m_u)^{D-1}+(\langle\sigma\rangle+m_d)^{D-1}}{2\langle\sigma\rangle+m_u+m_d},
\label{mass:pi}
\end{equation}
and
\begin{equation}
   Z_{\pi^{3}}^{-1}M_0^{D-4} m_{\pi^{3}}^2 = \frac{1}{2 g_\pi^0} 
   - \frac{1}{2}A(D) 
  \sum_{j\in \{u,d\}}[(m_j+\langle\sigma\rangle)^2]^{D/2-1}.
\label{mass:pi:3}
\end{equation}
In the massless quark limit Eqs.(\ref{mass:pi}) and (\ref{mass:pi:3}) read
\begin{equation}
   Z_{\pi}^{-1}M_0^{D-4} m_\pi^2 
   \rightarrow \frac{1}{2 g_\pi^0} - A(D) (\langle\sigma\rangle^2)^{D/2-1}.
\label{mass:pi2}
\end{equation}
Substituting the non-trivial solution of the gap equation 
(\ref{sol2:gap}) into Eq.(\ref{mass:pi2}), we observe that the pion 
mass disappears. Therefore the pion propagator has massless pole and thus the 
Nambu-Goldstone mode remains in the dimensional regularization.
This is 
natural given the fact that dimensional regularization does not 
damage Word-Takahashi identity, whereas cut-off regularization does 
\cite{AIBY}.  
If the constituent quark mass is not generated, the above expression
can be written as 
\begin{equation}
   Z_{\pi}^{-1}M_0^{D-4} m_\pi^2
   \rightarrow \frac{1}{2 g_\pi^0}
   = \left(\frac{1}{2g_\pi^r}-\frac{1}{2g_\pi^{cr}}\right)(M_0^2)^{D/2-1}.
\label{mass:pi3}
\end{equation}
Since the renormalization constant $Z_\pi$ is positive definite, the pion 
mass squared is negative in the broken phase, $1/g_\pi^r < 1/g_\pi^{cr}$. 
Such a tachyon pole is compensated by the constituent quark mass. On
the other hand 
inverse of the renormalization constant $Z_\pi^{-1}$ is divergent
and a massless pion pole appears in the symmetric phase, 
$1/g_\pi^r > 1/g_\pi^{cr}$.

\subsection{The scalar meson}  

Next we calculate the meson mass 
in the scalar channel. The propagator for 
the scalar channel 
 in the leading order of the $1/N$ expansion is given in Eq.(\ref{112})
\begin{equation}
  G_{s}(p^2,\langle\sigma\rangle) 
  = \frac{4(g_\pi^0)^2}{2 g_\pi^0 - \Pi_s(p^2)} .
\label{pro:s}
\end{equation}
The scalar meson mass (given that the corresponding pole is not located on the real axis) is obtained by observing the extremum for
$|G_{s}(p^2=m_\sigma^2,\langle\sigma\rangle)|$.
We can find the corresponding pole for $D \gtrsim 2.2$. It can not be
found for $D \lesssim 2.2$ and for soft mode in the next section.
The extremum condition Eq.~(\ref{pro:s}) for scalar meson mass is
applicable for more general cases.

First we discuss the massless pole in the scalar channel. It is not
realistic but useful to understand the symmetry properties of NJL model.
If the scalar meson is light enough, we can employ the same procedure 
as
in the previous subsection.
Thus we 
use the definition for $m_\sigma$ similar to $m_\pi$. 
\begin{equation}
  G_{s}(p^2,\langle\sigma\rangle)
  \equiv \frac{Z_\sigma M_0^{4-D}}{m^2_{\sigma}-p^2+O(p^4)}.
\label{pro:s2}
\end{equation}
The self-energy $\Pi_s(p^2)$ defined in Eq.(\ref{pis}) 
is expanded near $p^2\sim 0$ as
\begin{eqnarray}
  \Pi_s(p^2\sim 0) &=& 2 (g^0_\pi)^2 (D-1) A(D)
  \left[ \sum_{j\in \{u,d\}}[(m_j+\langle\sigma\rangle)^2]^{D/2-1} \right.
\nonumber \\ && \left.
   +\frac{1}{6}\left(1-\frac{D}{2}\right)p^2
   \sum_{j\in \{u,d\}}[(m_j+\langle\sigma\rangle)^2]^{D/2-2}
  \right] +O(p^4).
\end{eqnarray}
Thus the renormalization constant for the scalar wave function is
obtained by
\begin{eqnarray}
  Z_\sigma^{-1} &=& \frac{D-1}{12}
   \left(1-\frac{D}{2}\right)A(D)M_0^{4-D}
   \sum_{j\in \{u,d\}}[(m_j+\langle\sigma\rangle)^2]^{D/2-2} .
\end{eqnarray}
It has a positive definite value for $2<D<4$. In the present approximation
the scalar mass is given by
\begin{equation}
   Z_\sigma^{-1}M_0^{D-4}m_\sigma^2 
   = \frac{1}{2 g_\pi^0} - \frac{1}{2} A(D)(D-1) 
   \sum_{j\in \{u,d\}}[(m_j+\langle\sigma\rangle)^2]^{D/2-1}.
\label{mass:sigma}
\end{equation}
In the massless quark limit it becomes
\begin{equation}
   Z_\sigma^{-1}M_0^{D-4}m_\sigma^2 \rightarrow \frac{1}{2 g_\pi^0} (2-D)
   =\left(\frac{1}{2g_\pi^r}-\frac{1}{2g_\pi^{cr}}\right)(2-D)(M_0^2)^{D/2-1}.
\label{mass:sigma0}
\end{equation}
where we used the gap equation (\ref{sol2:gap}). As is shown in Sec.II,
the bare coupling $g_\pi^0$ is always negative in the broken phase for 
$2<D<4$. Thus the sigma mass squared is real and positive
for $1/g_\pi^r > 1/g_\pi^{cr}$. 
The sigma mass decreases as the inverse of the coupling $1/g_\pi^r$ increases.
For $1/g_\pi^{r} \ge 1/g_\pi^{cr}$ the renormalization
constant $Z_\sigma$ vanishes. Hence a massless pole appears for the
scalar channel. The sigma meson degenerates with the pion in the
symmetric phase.

If we do not consider the constituent quark mass, 
the sigma meson
mass degenerates with the pion mass (\ref{mass:pi3}) in the limit of massless
 quarks. The tachyon pole (\ref{mass:pi3}) shows that the 
vacuum is unstable and the constituent quark mass should be generated. 

Since the scalar meson is heavy in the real world, we do not apply
the expression (\ref{mass:sigma}) to evaluate the real scalar meson 
mass. Furthermore, there is no pole for 
$p^2\leq4(\langle\sigma\rangle+m)^2$. 
An imaginary part appears in the self-energy for 
$p^2>4(\langle\sigma\rangle+m)^2$.
The imaginary part corresponds to the decay width of the sigma meson.
To find the sigma meson resonance we numerically calculate the pole 
of the propagator.

\subsection{Dimensions and physical scale}
In this subsection we discuss the physical scale and coupling constant 
in terms of the space-time dimensions. Since the NJL model is phenomenological 
model of QCD, the scale of the model should be determined to reproduce the 
observable physical quantities. Here we calculate the pion decay constant
and the pion mass as typical quantities in the hadronic phase.
To consider the QCD interaction we set $N_c=3$. 
In the present model the space-time dimensions $D$ 
in the loop integrals is also regarded as one 
of the parameters of the phenomenological model.

Inserting the measured values $m_\pi^0=135.0$MeV and $m_\pi^\pm=139.6$MeV 
into Eqs.(\ref{mass:pi}) and (\ref{mass:pi:3}), we can obtain the matching 
condition for the coupling constant $g_\pi$ and the space-time dimensions $D$ 
with the physical scale. But 
the main part of the mass difference between the neutral 
and the charged pion arises from the electro-magnetic interaction which is not 
considered here \cite{DSTL}. 
Thus we inspect only the isospin symmetric 
limit $m_\pi^0=m_\pi^\pm\simeq 136$Mev and set $m_u=m_d$ below.

We start with the relationship in the current algebra (See, for example, 
appendix in Ref. \cite{IKM2}.) The pion decay constant $f_\pi$ is written as
\begin{equation}
  f_{\pi} = -Z_{\pi}^{-1/2}\langle\sigma\rangle.
\label{decay}
\end{equation}
From the Gell-Mann-Oakes-Renner relation \cite{GOR} we obtain 
\begin{equation}
  f_{\pi}^2 m_{\pi}^2 = \frac{\langle\sigma\rangle}{4 g_\pi^0} {M_0}^{4-D} 
  (m_u + m_d) .
\label{decay2}
\end{equation}
The constituent quark mass $\langle\sigma\rangle$ is determined by the gap equation 
(\ref{gap}). Substituting the measured values $m_\pi = 136$MeV, 
$f_\pi = 93$MeV and the solution of the gap equation we obtain
the matching condition for the coupling constant $g_\pi$, the space-time 
dimensions $D$ and the renormalization scale $M_0$ with the physical 
scale. After some numerical calculations we get the physical scale for
$g_\pi$ and $M_0$ as a function of the space-time dimensions $D$.

\begin{figure}[t]
\begin{minipage}{66mm}
\begin{center}
\includegraphics[width=66mm]{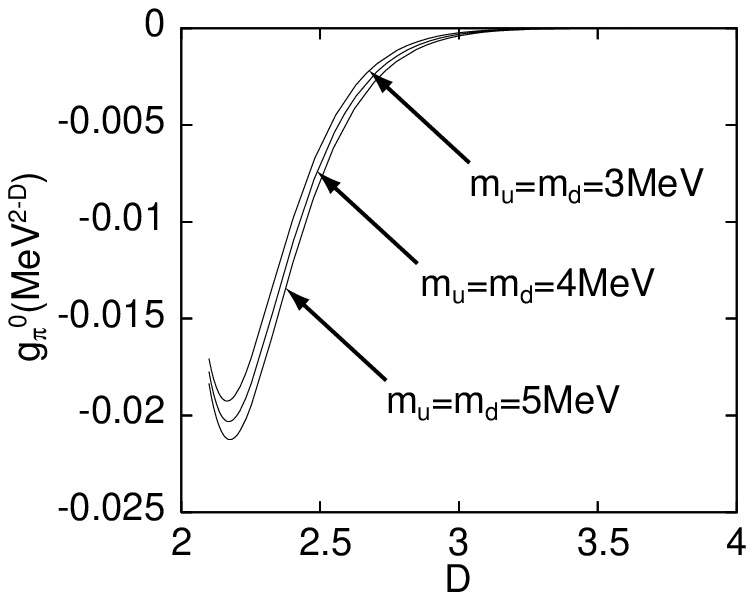}
\end{center}
\end{minipage}
\begin{minipage}{66mm}
\begin{center}
\includegraphics[width=63mm]{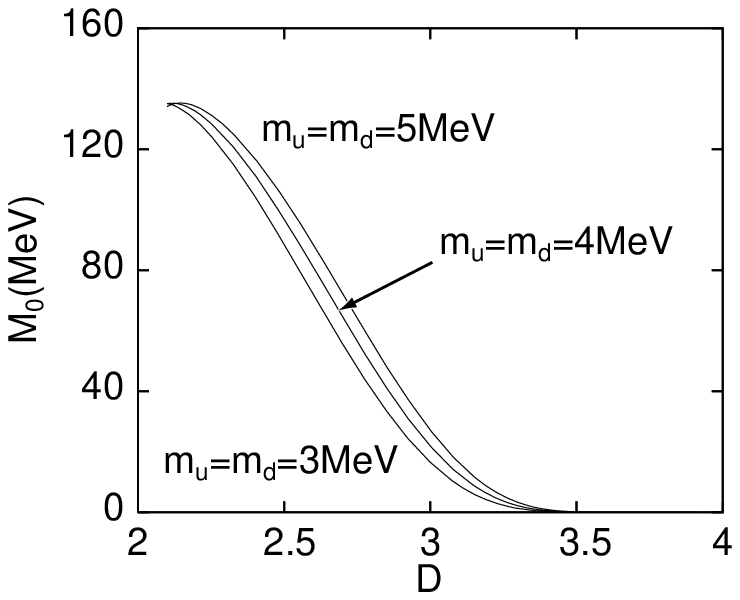}
\end{center}
\end{minipage}
\caption{The physical scale for the bare coupling constant $g_\pi^0$ 
and the renormalization scale $M_0$ are shown as a function of the space-time
dimensions $D$ for $N_c=3$, $m_u=m_d=3$, $4$ and $5$MeV.}
\label{fig:D}
\end{figure}

\begin{figure}[t]
\begin{minipage}{66mm}
\begin{center}
\includegraphics[width=65mm]{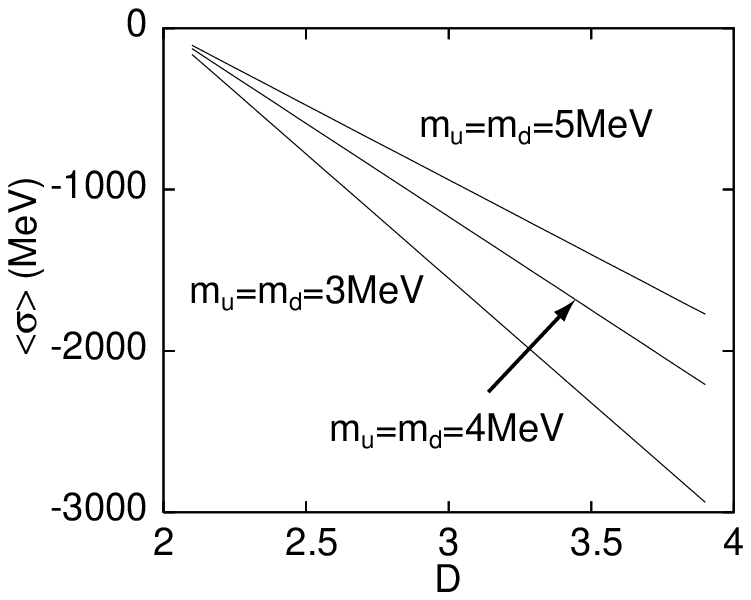}
\end{center}
\end{minipage}
\begin{minipage}{66mm}
\begin{center}
\includegraphics[width=66mm]{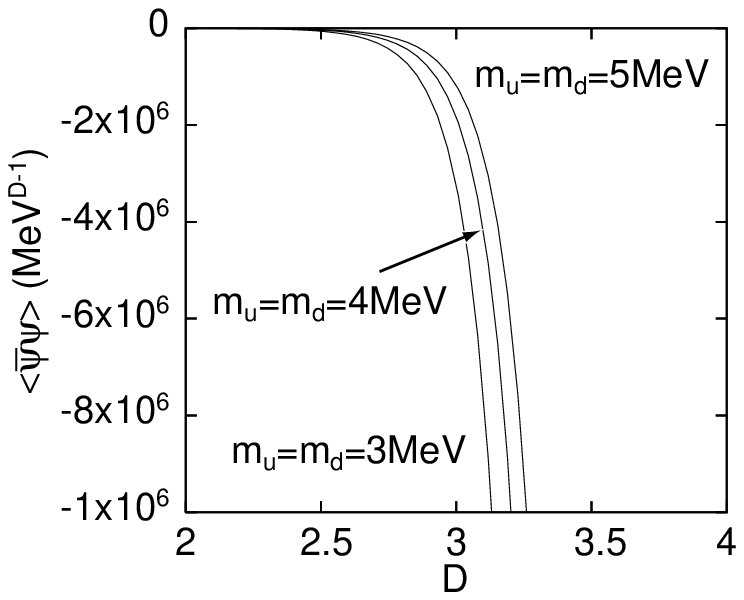}
\end{center}
\end{minipage}
\caption{The constituent quark mass $\langle\sigma\rangle$ and the expectation value
$\langle\bar\psi\psi\rangle$ are shown as a function of the space-time
dimensions $D$.}
\label{fig:pi}
\end{figure}

In Fig.2  the curves  satisfy the relations (\ref{decay}) 
and (\ref{decay2}) on $D-g_\pi^0$ and $D-M_0$ 
plane for $m_u=m_d=3$, $4$ 
and $5$MeV. As is shown in Fig. 2, the bare coupling constant and the
renormalization scale has no 
strong dependence on the current quark mass.
The renormalization scale is found to be below the pion mass. In Fig.3
we illustrate the behavior of the constituent quark mass and 
of the expect value of the composite operator of a quark and an anti-quark, 
$\displaystyle \langle\bar\psi\psi\rangle\sim -\frac{1}{2 g_\pi^0} \langle\sigma\rangle$, 
for the parameters fixed by (\ref{decay}) and (\ref{decay2}).
To obtain the constituent quark mass near $300$MeV we should consider 
the lower space-time dimensions or larger current quark mass. 
The expectation value $\langle\bar\psi\psi\rangle$ has 
strong dependence 
on the space-time dimensions. But the normalized
value, $\langle\bar\psi\psi\rangle M_0^{4-D}$, varies less than 
$0.1\%$ between two and four dimensions and is found to be
\begin{equation}
\langle\bar{u}u\rangle M_0^{4-D} \simeq
\left\{
  \begin{array}{ll}
    -(298.7 \mbox{MeV})^3 & (m_u=m_d=3\mbox{MeV}),\\
    -(271.4 \mbox{MeV})^3 & (m_u=m_d=4\mbox{MeV}),\\
    -(252.0 \mbox{MeV})^3 & (m_u=m_d=5\mbox{MeV}).
  \end{array}
\right.
\end{equation}

Inserting the non-trivial solution of the gap equation into 
Eq.(\ref{mass:pi:3}) and taking $m_u=m_d=m$ we get
\begin{equation}
  Z_\pi^{-1}M_0^{D-4}m_\pi^2 \simeq \frac{m}{2g_\pi^0 \langle\sigma\rangle}+\mbox{O}
  \left(\left(\frac{m}{\langle\sigma\rangle}
  \right)^2\right) .
\label{new:pi}
\end{equation}
The Gell-Mann-Oakes-Renner relation (\ref{decay2}) is reproduced from 
Eqs.(\ref{decay}) and (\ref{new:pi}).
Thus the Eq. (\ref{mass:pi:3}) should be consistent with the model 
independent relationships (\ref{decay}) and (\ref{decay2}).

\begin{figure}[t]
\begin{minipage}{66mm}
\begin{center}
\includegraphics[width=66mm]{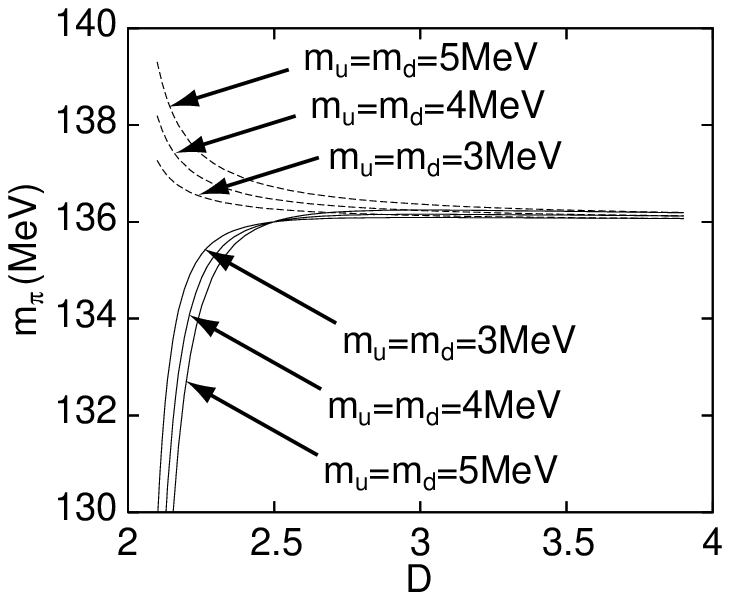}

(a) Pion mass from Eq.(\ref{mass:pi:3}) (dashed line) 
and Eq.(\ref{pi:mass:num}) (full line).
\end{center}
\end{minipage}
\begin{minipage}{66mm}
\begin{center}
\includegraphics[width=66mm]{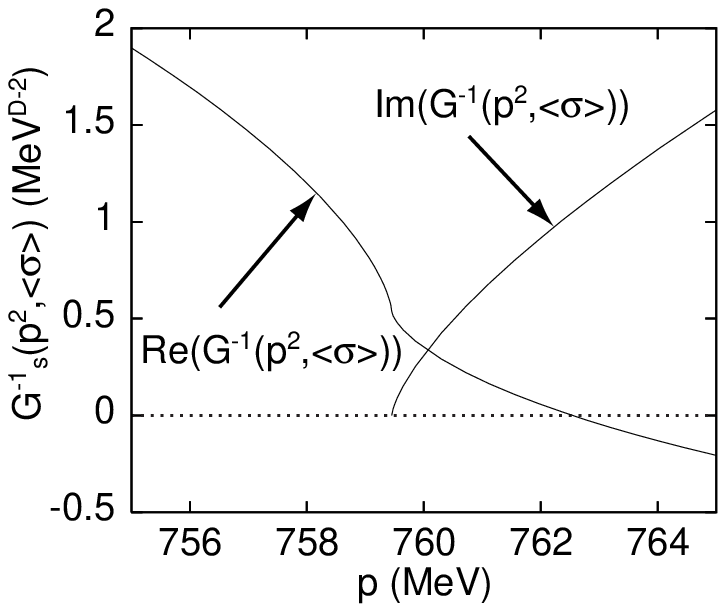}
\noindent

(b) $G^{-1}_s(p^2,\langle\sigma\rangle)$ for $D=2.4$ and $m_u=m_d=5$MeV.
\end{center}
\end{minipage}
\caption{Solutions for Eq.(\ref{pi:mass:num}) and Eq.(\ref{sigma:mass:num}).}
\label{fig:mpi}
\end{figure}

\begin{figure}[t]
\begin{minipage}{66mm}
\begin{center}
\vglue 6mm
 \includegraphics[width=66mm,height=55mm]{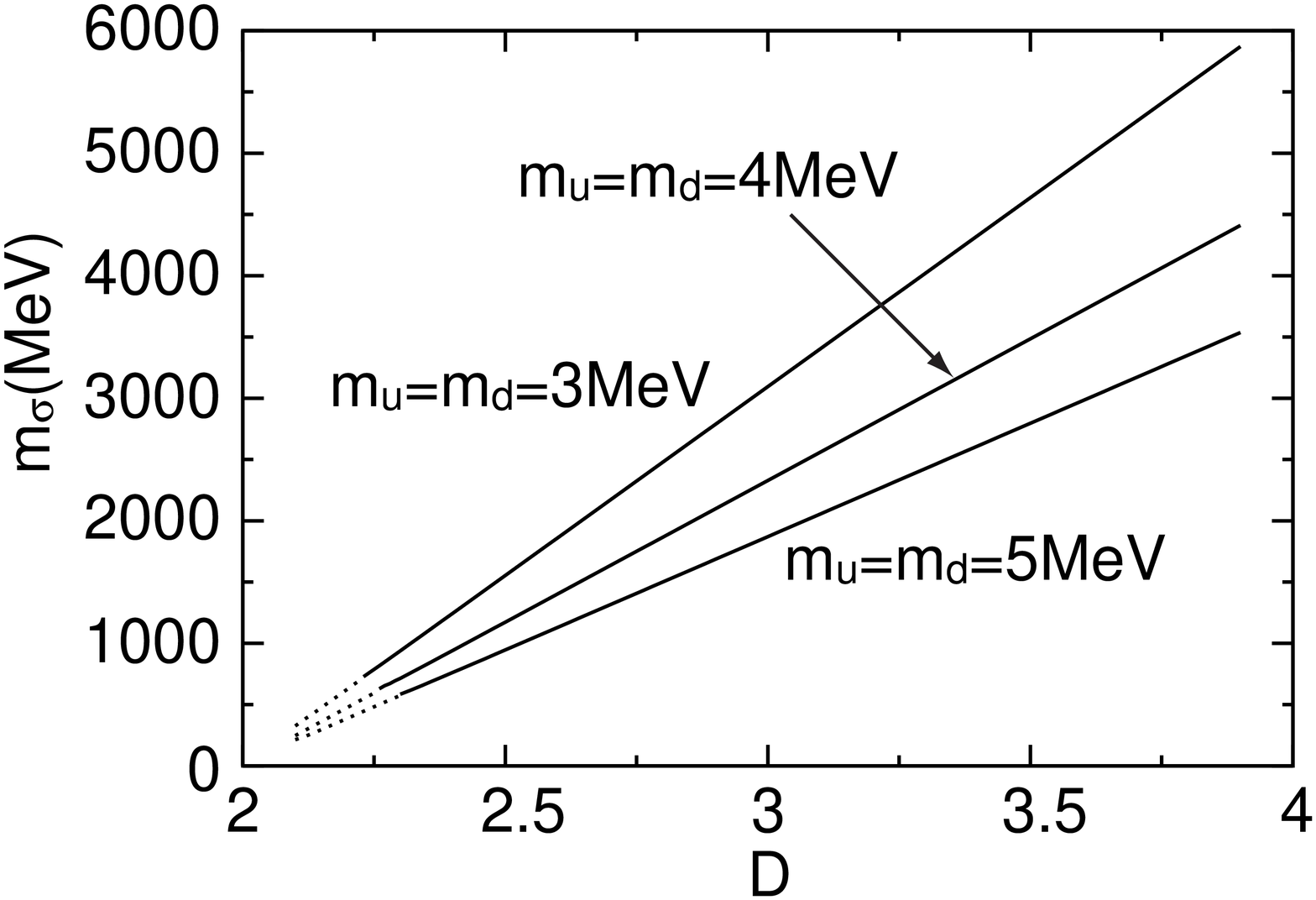}

(a) Sigma meson mass from Eq. (\ref{sigma:mass:num}) (full line)
   and Eq. (\ref{sigma:mass:num2}) (dotted line).
\end{center}
\end{minipage}
\begin{minipage}{66mm}
\begin{center}
 \includegraphics[width=66mm,height=55mm]{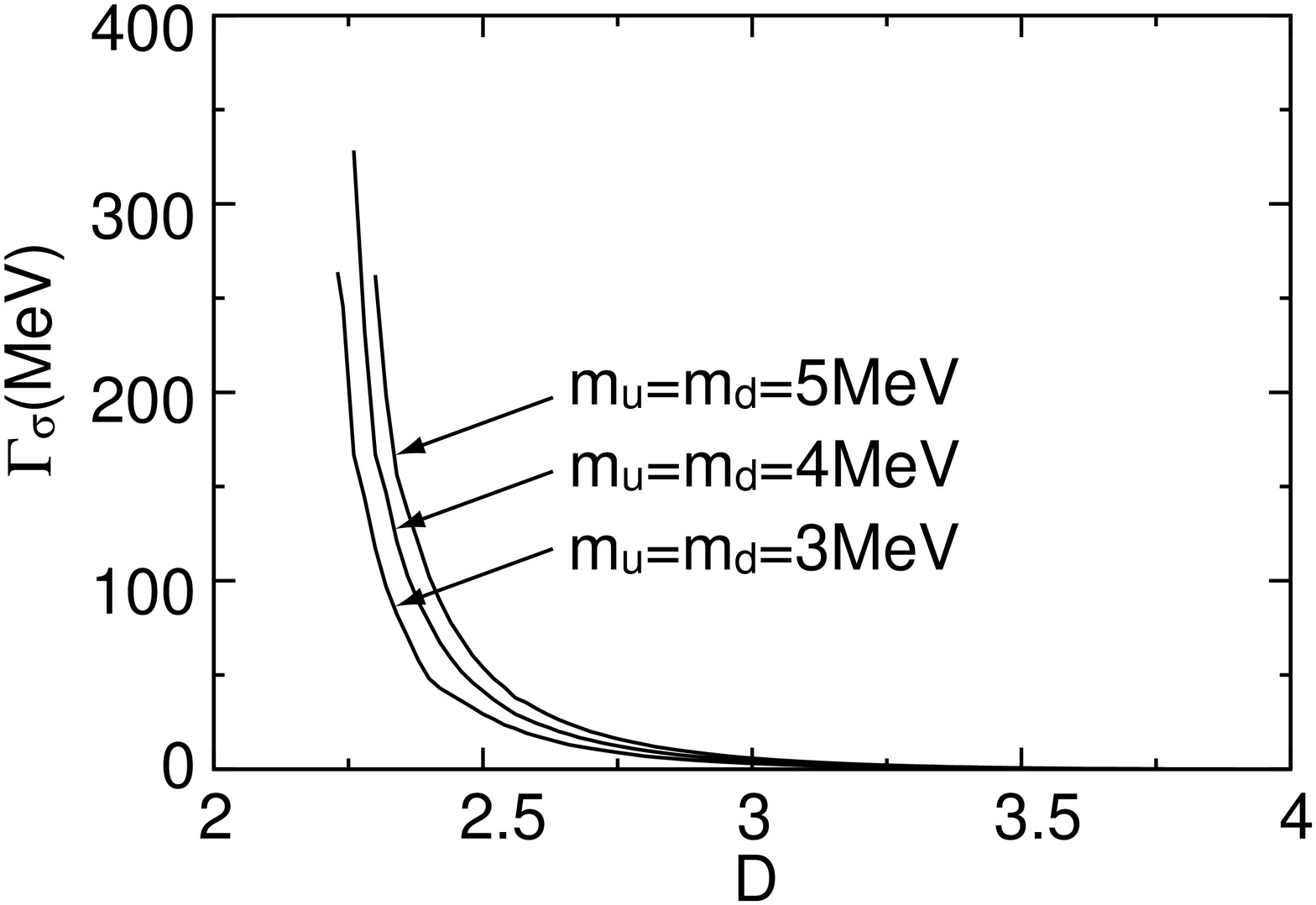}
\noindent

(b) Sigma meson width from Eq. (\ref{sigma:width}).
\end{center}
\end{minipage}
\caption{Sigma meson mass and width}
\label{fig:msigma}
\end{figure}

In Figs.4 and 5
\footnote{We choose a space-time dimensions at $D=2.4$ in 
Fig.~\ref{fig:mpi} (b) and so on. The sigma meson mass for $D=2.4$ seems
to be a little bit larger than the real value, but general properties 
for the phase structure are similar
to the other space-time dimensions.}
we calculate the pion and the sigma meson mass by 
numerically solving the equations
\begin{equation}
  G^{-1}_5(p^2=m_\pi^2, \langle\sigma\rangle) 
  = \frac{1}{2g^0_\pi}-\frac{1}{4 (g^0_\pi)^2}
  \Pi_5(p^2=m_\pi^2)=0,
\label{pi:mass:num}
\end{equation}
and 
\begin{equation}
  G^{-1}_s(p^2=m_\sigma^2, \langle\sigma\rangle) 
  = \frac{1}{2g^0_\pi}-\frac{1}{4 (g^0_\pi)^2}
    {\rm Re}\left[ \Pi_s(p^2=m_\sigma^2) \right]=0,
\label{sigma:mass:num}
\end{equation}
using the gap equation. 
As is clearly seen in Fig.~4 (a), the results for $m_\pi$ are found 
at the physical pion mass scale $\simeq 136$MeV for $D \gtrsim 2.2$. 
The approximation in Eq. (\ref{mass:pi:3}) seems to be valid for 
higher dimensions.
A typical behavior of the real and the imaginary part of 
$G^{-1}_s(p^2, \langle\sigma\rangle) $ is shown in Fig.~4 (b).
Since the sigma meson is almost stable in the
leading order of the $1/N$ expansion. We use the approximate equation
(\ref{sigma:mass:num}). The approximation is not valid in some cases,
for example, $D \lesssim 2.3$, hight temperature and/or large chemical
potential. In such cases we use
\begin{equation}
  \frac{\partial}
  {\partial p} |G_s(p^2=m_\sigma^2, \langle\sigma\rangle)| = 0 .
\label{sigma:mass:num2}
\end{equation}
The sigma meson develops the heavier mass for higher dimensions in 
Fig.~5 (a). If imaginary part of $G_s$ is small, we can identify the
sigma meson width, $\Gamma_\sigma$, as
\begin{equation}
 \Gamma_\sigma = - \frac{Z_\sigma M_0^{4-D}}{4(g_\pi^0)^2 m_\sigma}
                 {\rm Im} \left[ \Pi_s(p^2=m_\sigma^2)  \right] .
\label{sigma:width}
\end{equation} 
with $m_\sigma$ obtained from Eq. (\ref{sigma:mass:num}).
We plot the behavior of the sigma meson width in Fig.~5 (b).
For smaller dimensions both the sigma meson and width approach to the
result obtained by solving the Roy equations in \cite{Capr}.
As is pointed out in Ref~\cite{Sann},
higher order corrections, $\pi\pi$ scattering, are essential for the 
sigma width. To discuss the sigma meson width one should
calculate the next to leading order corrections in $1/N$ expansion.

\section{Dynamical symmetry breaking at finite $T$ and $\mu$}

To define the thermal equilibrium the time direction has to
be fixed. 
This breaks general covariance 
however, it is important not to introduce an additional violation of the 
covariance through the regularization.
Thus we apply the dimensional regularization to the NJL model
in the thermal equilibrium and extend the above analysis to the finite 
temperature $T$ and the chemical potential $\mu$. 

First we discuss the phase structure of the theory.
It is expected that the broken chiral symmetry is restored at high
temperature and/or large chemical potential. 
Following the standard procedure of the imaginary-time formalism, 
we introduce the temperature and the chemical potential to the NJL model 
in $D$ dimensions \cite{IKM}.
Hence the gap equation (\ref{eq:gap}) is modified as
\begin{equation}
  \langle\sigma\rangle  =   2 g_\pi^0 \frac{1}{\beta} \sum_{n}
  \int \frac{d^{D-1} \boldsymbol{k}}{(2\pi)^{D-1}} \mbox{tr} S_{\beta\mu}(k),
\label{eq:gap:ft}
\end{equation}
where $S_{\beta\mu}(k)$ is the fermion propagator at finite $T$ and $\mu$,
\begin{equation}
  S_{\beta\mu}(k) \equiv \frac{1}{
  \boldsymbol{k} \cdot\boldsymbol{\gamma} -(\omega_n -i\mu) \gamma_4 
  +m+\langle\sigma\rangle-i\epsilon}.
\end{equation}
Due to the anti-periodicity of the fermion field the $k^0$ integral
is replaced by the summation over $\omega_n=(2n+1)\pi/\beta$.
We note that the flavor symmetry is broken by the quark
mass and the chemical potential in the thermal equilibrium
(see, for example, Ref.\cite{LB}.)

Integrating over the momentum in Eq.(\ref{eq:gap:ft}), we obtain
\begin{eqnarray}
  \langle\sigma\rangle&=&-\sqrt{2} g_\pi^0 A(D-1)\sum_{j\in \{u,d\}}(m_j+\langle\sigma\rangle)
\nonumber \\
&& \times \frac{1}{\beta}
  \sum_n\left[(\omega_n-i\mu)^2+(m_j+\langle\sigma\rangle)^2\right]^{(D-3)/2}.
\label{eq:gap:ft1}
\end{eqnarray}
We also perform the summation in Eq.(\ref{eq:gap:ft}) and get another 
expression,
\begin{eqnarray}
  \langle\sigma\rangle & = & g_\pi^0 A(D) \sum_{j\in \{u,d\}} 
  (m_j+\langle\sigma\rangle) [(m_j+\langle\sigma\rangle)^2]^{D/2-1}
\nonumber \\
  &&      - \frac{2\sqrt{2}N_c g_\pi^0}{(2\pi)^{(D-1)/2}}
            \frac{1}{\displaystyle \Gamma\left(\frac{D-1}{2}\right)}
          \int_0^\infty k^{D-2} dk \ 
          \sum_{j\in \{u,d\}} \frac{m_j+\langle\sigma\rangle}{\sqrt{k^2+(m_j+\langle\sigma\rangle)^2}}
\nonumber \\
  && \times \left(\frac{1}{1+e^{\beta(\sqrt{k^2+(m_j+\langle\sigma\rangle)^2}+\mu)}}
                + \frac{1}{1+e^{\beta(\sqrt{k^2+(m_j+\langle\sigma\rangle)^2}-\mu)}}
            \right) .
\label{eq:gap:ft2}
\end{eqnarray}
The result exactly reproduces the one obtained in Ref.\cite{IKM}
at the massless quark limit. 

At the zero temperature limit
the gap equation (\ref{eq:gap:ft2}) simplifies to
\begin{eqnarray}
  \langle\sigma\rangle & = & g_\pi^0 A(D) \sum_{j\in \{u,d\}} (m_j+\langle\sigma\rangle) [(m_j+\langle\sigma\rangle)^2]^{D/2-1}
\nonumber \\
  &&      - \frac{2\sqrt{2}N_c g_\pi^0}{(2\pi)^{(D-1)/2}}
            \frac{1}{\displaystyle \Gamma\left(\frac{D-1}{2}\right)}
          \sum_{j\in \{u,d\}} 
          \int_0^\infty k^{D-2} dk \frac{m_j+\langle\sigma\rangle}{\sqrt{k^2+(m_j+\langle\sigma\rangle)^2}}
\nonumber \\
  && \times \theta (\mu - \sqrt{k^2+(m_j+\langle\sigma\rangle)^2}) .
\label{eq:gap:ft3}
\end{eqnarray}
For $\mu < m_u+\langle\sigma\rangle$ and $\mu < m_d+\langle\sigma\rangle$ the theta function vanishes and
we obtain the same solution with zero chemical potential.
\begin{equation}
  \langle\sigma\rangle = g_\pi^0 A(D) \sum_{j\in \{u,d\}} (m_j+\langle\sigma\rangle) [(m_j+\langle\sigma\rangle)^2]^{D/2-1}.
\end{equation}
Hence, the chemical potential does not contribute to the constituent
quark mass $\langle\sigma\rangle$ at $T=0$, $\mu < m_u+\langle\sigma\rangle$ and $\mu < m_d+\langle\sigma\rangle$.

To find the stable solution we evaluate the effective potential $V(\sigma)$ 
at finite $T$ and $\mu$.
It is obtained by integrating over $\langle \sigma \rangle$ 
in the gap equation (\ref{eq:gap:ft2}) with further replacement 
$\langle \sigma \rangle$ by $\sigma$
\begin{eqnarray}
  V(\sigma) & = & \frac{\sigma^2}{4g_\pi^0}
  - \frac{A(D)}{2D} \sum_{j\in \{u,d\}} [(m_j+\sigma)^2]^{D/2}
\nonumber \\
  &&      - \frac{1}{\beta}\frac{\sqrt{2}N_c}{(2\pi)^{(D-1)/2}}
            \frac{1}{\displaystyle \Gamma\left(\frac{D-1}{2}\right)}
          \int_0^\infty k^{D-2} dk \ 
          \sum_{j\in \{u,d\}}
\label{eq:epot:ft}
\\
  && \times \left(\ln\frac{1+e^{-\beta(\sqrt{k^2+(m_j+\sigma)^2}+\mu)}}
                          {1+e^{-\beta(\sqrt{k^2+m_j^2}+\mu)}}
                + \ln\frac{1+e^{-\beta(\sqrt{k^2+(m_j+\sigma)^2}-\mu)}}
                          {1+e^{-\beta(\sqrt{k^2+m_j^2}-\mu)}}
            \right) .
\nonumber
\end{eqnarray}
\begin{figure}[t]
\begin{minipage}{66mm}
\begin{center}
\vglue 2mm
\includegraphics[width=64mm]{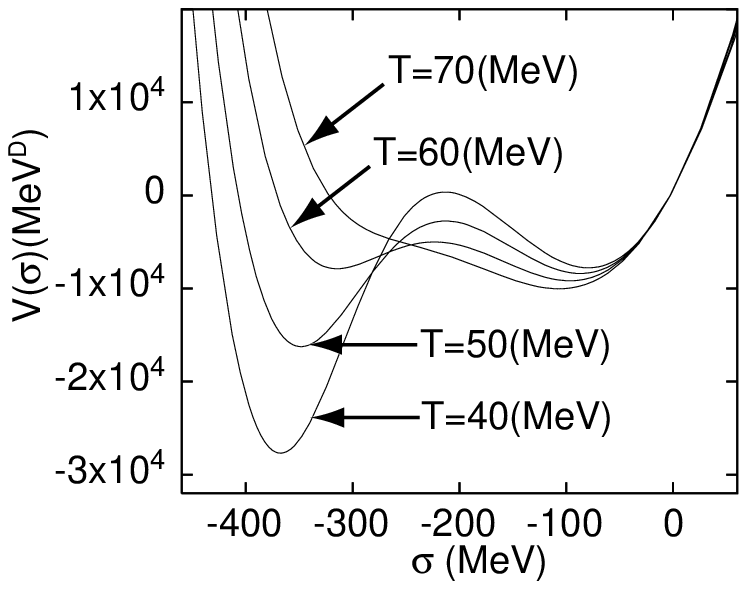}

(a) Behavior of the effective potential at $\mu=300$(MeV). 
\end{center}
\end{minipage}
\begin{minipage}{66mm}
\begin{center}
\includegraphics[width=64mm]{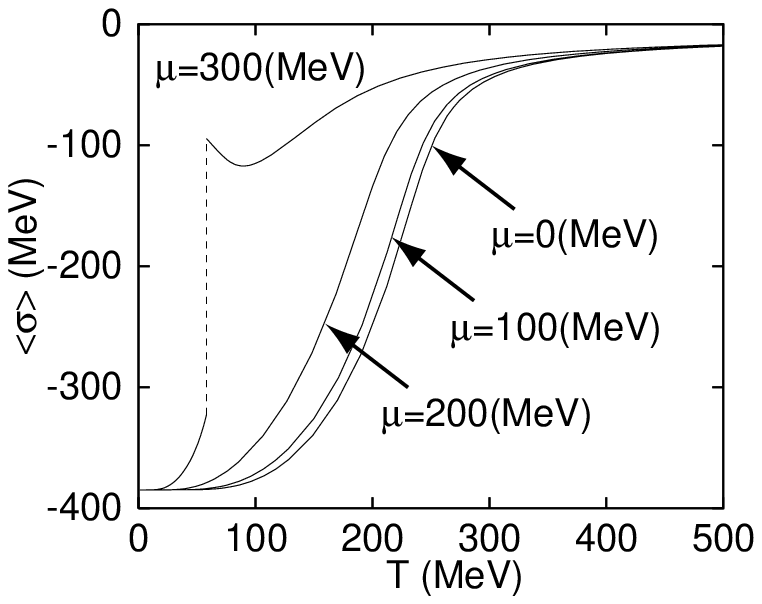}

(b) Behavior of the constituent quark mass $\langle\sigma\rangle$.
\end{center}
\end{minipage}
\caption{Typical behavior of the effective potential and the constituent 
quark mass at finite $T$ and $\mu$ in $D=2.4$
and $m_u=m_d=5$MeV.}
\label{fig:potft}
\end{figure}
The stable solution of the gap equation is obtained by observing the
minimum of the effective potential. We numerically calculate it and show
typical behavior of the effective potential $V(\langle\sigma\rangle)$ 
and of the dynamically 
generated quark mass $\langle\sigma\rangle$ in Fig.\ref{fig:potft}.
A first order transition takes place
for $\mu=300$(MeV). In the massless quark limit a phase transition occurs
as is shown in Ref.\cite{IKM}. Because of the explicit 
chiral symmetry 
breaking term, i.e.
finite current quark mass, the phase transition becomes a cross over in
Fig.\ref{fig:potft}  \cite{HK2, K}.
In Ref.\cite{IKM} detailed behavior of the dynamically generated mass 
and the phase structure are found for $m_u=m_d=0$. Analytic expressions 
for some characteristic points are also presented there.

\section{Meson masses at finite $T$ and $\mu$}

Next we consider 
meson masses in a thermal equilibrium.
The question of whether the Nambu-Goldstone modes of the chiral
symmetry breaking are still massless or not at finite $T$ and 
$\mu$ is not trivial
 \cite{IM,PT,T,P,LR}. It is expected that the dispersion
law is modified in the thermal equilibrium.

In the imaginary time formalism the self-energy for the pseudo-scalar 
channel, $\Pi_{5}^{a}(p^2)$ is modified as 
\begin{equation}
  \Pi_{5}^{a}(p_4, \boldsymbol{p}) = 
  - 4 (g_\pi^0)^2 \frac{1}{\beta} \sum_{n}
  \int \frac{d^{D-1} \boldsymbol{k}}{(2\pi)^{D-1}} \mbox{tr} 
  [i\gamma^5 \tau^a S(k)_{\beta\mu} i\gamma^5 \tau^a S(k-p)_{\beta\mu}] .
\label{sel:pi:ft}
\end{equation}
Performing the trace operation in Eq.(\ref{sel:pi:ft}), we obtain
\begin{eqnarray}
\Pi_{5}^{1,2}(p_4, \boldsymbol{p}) &=& 
  4 N_c  (g_\pi^0)^2 2^{D/2} \frac{1}{\beta} 
  \sum_{n} \int \frac{d^{D-1} \boldsymbol{k}}{(2\pi)^{D-1}}
\nonumber \\
&&\hspace{-10mm}\times 
  \frac{(\omega_n-i\mu)(\omega_n-p_4-i\mu)+\boldsymbol{k}\cdot(\boldsymbol{k-p})+(m_u+\langle\sigma\rangle)(m_d+\langle\sigma\rangle)}
  {[(\omega_n-i\mu)^2+{E_{1u}}^2][(\omega_n-p_4-i\mu)^2+{E_{2d}}^2]}
\nonumber \\[2mm]
&&\hspace{-10mm}+(u\leftrightarrow d),
\label{sel:pi0:ftp}
\end{eqnarray}
and
\begin{eqnarray}
  \Pi_{5}^{3}(p_4, \boldsymbol{p}) &=& 
   4 N_c  (g_\pi^0)^2 2^{D/2} \frac{1}{\beta} 
  \sum_{n} \int \frac{d^{D-1} \boldsymbol{k}}{(2\pi)^{D-1}} \sum_{j\in\{u,d\}}
\label{sel:pi0:ftp:3}
\\ \nonumber
  &&\times 
  \frac{(\omega_n-i\mu)(\omega_n-p_4-i\mu)+\boldsymbol{k}\cdot(\boldsymbol{k-p})+(m_j+\langle\sigma\rangle)^2}
  {[(\omega_n-i\mu)^2+{E_{1k}}^2][(\omega_n-p_4-i\mu)^2+{E_{2k}}^2]} ,
\end{eqnarray}
where
\begin{equation}
  {E_{1k}}^2=\boldsymbol{k}^2+(m_j+\langle\sigma\rangle)^2,\ \ \ 
  {E_{2k}}^2=(\boldsymbol{k-p})^2+(m_j+\langle\sigma\rangle)^2 .
\end{equation}
We observe the massless pole of the pion propagator at the massless quark 
limit, again.
For $m_u=m_d=0$ Eqs.(\ref{sel:pi0:ftp}) and (\ref{sel:pi0:ftp:3}) reduce to
\begin{eqnarray}
  \Pi_{5}^{a}(p_4, \boldsymbol{p}) &=& 
   4 N_c  (g_\pi^0)^2 2^{D/2} \frac{1}{\beta} 
  \sum_{n} \int \frac{d^{D-1} \boldsymbol{k}}{(2\pi)^{D-1}}
\nonumber
\\ 
  &&\times 
  \left[ \frac{1}
  {(\omega_n-i\mu)^2+{E_1|_{m=0}}^2}
  + \frac{1}
  {(\omega_n-p_4-i\mu)^2+{E_2|_{m=0}}^2}
\right.
\nonumber \\
  && \left.
  - \frac{p_4^2+\boldsymbol{p}^2}
  {[(\omega_n-i\mu)^2+{E_1|_{m=0}}^2][(\omega_n-p_4-i\mu)^2+{E_2|_{m=0}}^2]}\right] .
\label{sel:pi0:ftp:a}
\end{eqnarray}
To find the massless pole we put $p^2=p_4^2+\boldsymbol{p}^2=0$.
Then the final line in Eq.(\ref{sel:pi0:ftp:a}) disappears.
Since the pseudo-scalar field with the momentum $p$ obeys 
the Bose-Einstein statistics, the fourth element 
$p_4$ is written as $p_4=2l\pi/\beta$ ($l$ is an integer).
We can simply change variables $\omega_n-p_4$ and $\boldsymbol{k-p}$
in the second line of Eq.(\ref{sel:pi0:ftp:a}) to $\omega_n$ and 
$\boldsymbol{k}$ respectively.

The pion self-energy (\ref{sel:pi0:ftp:a}) reduces to
\begin{eqnarray}
  \Pi_{5}^{a}(p^2=0) &=& 
  8 N_c  (g_\pi^0)^2 2^{D/2} \frac{1}{\beta} 
  \sum_{n} \int \frac{d^{D-1} \boldsymbol{k}}{(2\pi)^{D-1}}
  \frac{1}
  {(\omega_n-i\mu)^2+\boldsymbol{k}^2+\langle\sigma\rangle^2}
\nonumber \\
   &=& -\frac{4(g_\pi^0)^2}{\langle\sigma\rangle}\frac{1}{\beta} \sum_{n}
  \int \frac{d^{D-1} \boldsymbol{k}}{(2\pi)^{D-1}} \mbox{tr} S_{\beta\mu}(k).
\label{sel:pimassless:ft}
\end{eqnarray}
The pion mass at finite $T$ and $\mu$ is found to be
\begin{eqnarray}
   Z^{-1}_\pi m_\pi^2(p^2=0)M_0^{D-4} 
   &=& \frac{1}{2g_\pi^0}
   -\frac{\Pi_{5}^{a}(p^2=0)}{4(g_\pi^0)^2}
\nonumber \\
   &=& \frac{1}{2g_\pi^0}+ \frac{1}{\langle\sigma\rangle}\frac{1}{\beta} \sum_{n}
  \int \frac{d^{D-1} \boldsymbol{k}}{(2\pi)^{D-1}} \mbox{tr} S_{\beta\mu}(k).
\label{mass:pi:ftfm}
\end{eqnarray}
Substituting the gap equation (\ref{eq:gap:ft}) into Eq.(\ref{mass:pi:ftfm}),
we find that the pion is also massless at finite $T$ and $\mu$. The 
pion 
is still  massless if the quark state is represented by a non-trivial 
solution of the gap equation.
It is noteworthy that we do not use the Feynman parametrization,
it is not applicable to calculate loops at finite $T$ and $\mu$.

Next we evaluate the contribution of the finite current quark mass.
Here we consider the isospin symmetric limit for simplicity and set
$m_u=m_d\equiv m$ again. In this case a pole of the pion propagator is
given by
\begin{eqnarray}
   0 &=& \frac{1}{2g_\pi^0}
   -\frac{\Pi_{5}^{a}(p_4, \boldsymbol{p})}{4(g_\pi^0)^2}
\nonumber \\
   &=& \frac{1}{2g_\pi^0} \left(1-\frac{\langle\sigma\rangle}{\langle\sigma\rangle+m}\right)
       + N_c 2^{D/2} \frac{1}{\beta} 
       \sum_{n} \int \frac{d^{D-1} \boldsymbol{k}}{(2\pi)^{D-1}}
\nonumber \\ 
  &&\times 
  \frac{p_4^2+\boldsymbol{p}^2}
  {[(\omega_n-i\mu)^2+{E_1}^2][(\omega_n-p_4-i\mu)^2+{E_2}^2]} ,
\label{pole:pi:ftfm}
\end{eqnarray}
where we have used the gap equation (\ref{eq:gap:ft}) in passing from
the first line to the second and third
one. Performing the summation
in Eq.(\ref{pole:pi:ftfm}) \cite{LB}, we get
\begin{eqnarray}
   0 &=& \frac{1}{2g_\pi^0} \left(1-\frac{\langle\sigma\rangle}{\langle\sigma\rangle+m}\right)
       -  \frac{N_c 2^{D/2}(p_4^2+\boldsymbol{p}^2)}{4} 
          \int \frac{d^{D-1} \boldsymbol{k}}{(2\pi)^{D-1}}
          \frac{1}{E_1E_2}
\nonumber \\ 
  &&\times
  \left[\frac{1-n_+(E_1)-n_-(E_2)}{ip_4-E_1-E_2}
       -\frac{n_-(E_1)-n_-(E_2)}{ip_4+E_1-E_2}
\right. \nonumber \\ &&\left.  
       +\frac{n_+(E_1)-n_+(E_2)}{ip_4-E_1+E_2}
       -\frac{1-n_-(E_1)-n_+(E_2)}{ip_4+E_1+E_2}
  \right] ,
\label{pole:pi:ftfm:2}
\end{eqnarray}
where
\begin{equation}
  n_\pm(E)\equiv \frac{1}{e^{\beta(E\mp\mu)}+1} .
\end{equation}
To find the dispersion law in the Minkowski space we perform 
Wick 
rotation, $ip_4 \rightarrow p_0$. Thus the Eq.(\ref{pole:pi:ftfm:2}) reads
\begin{eqnarray}
   0 &=& \frac{1}{2g_\pi^0} \left(1-\frac{\langle\sigma\rangle}{\langle\sigma\rangle+m}\right)
       +  \frac{N_c 2^{D/2}(p_0^2-\boldsymbol{p}^2)}{4} 
          \int \frac{d^{D-1} \boldsymbol{k}}{(2\pi)^{D-1}}
          \frac{1}{E_1E_2}
\nonumber \\ 
  &&\times
  \left[\frac{1-n_+(E_1)-n_-(E_2)}{p_0-E_1-E_2}
       -\frac{n_-(E_1)-n_-(E_2)}{p_0+E_1-E_2}
\right. \nonumber \\ &&\left.  
       +\frac{n_+(E_1)-n_+(E_2)}{p_0-E_1+E_2}
       -\frac{1-n_-(E_1)-n_+(E_2)}{p_0+E_1+E_2}
  \right] .
\label{pole:pi:ftfm:3}
\end{eqnarray}
If we change variable $\boldsymbol{k}$ to $\boldsymbol{p-k}$
in Eq.(\ref{pole:pi:ftfm:3}), $E_1$ and $E_2$ are exchanged. 
Thus Eq.(\ref{pole:pi:ftfm:3}) simplifies to 
\begin{eqnarray}
   0 &=& \frac{1}{2g_\pi^0} \left(1-\frac{\langle\sigma\rangle}{\langle\sigma\rangle+m}\right)
       +  \frac{N_c 2^{D/2}(p_0^2-\boldsymbol{p}^2)}{2} 
          \int \frac{d^{D-1} \boldsymbol{k}}{(2\pi)^{D-1}}
\nonumber \\ 
  &&\times
  \left[\left(\frac{1}{E_1}+\frac{1}{E_2}\right)
\right. 
  \frac{1-n_+(E_1)-n_-(E_1)}
       {p_0^2-(E_1+E_2)^2}
\nonumber \\ &&
  +\left(\frac{1}{E_1}-\frac{1}{E_2}\right)
\left.
  \frac{-n_+(E_1)-n_-(E_1)}
       {p_0^2-(E_1-E_2)^2}
  \right] .
\label{pole:pi:ftfm:4}
\end{eqnarray}
We divide the momentum integral in the radial and the angle parts,
\begin{equation}
\int d^{D-1} \boldsymbol{k} = \int_0^{\infty}  k^{D-2} dk \int^{\pi}_0
\sin^{D-3}\theta_1 d\theta_1 \cdots \int^{\pi}_0 
\sin \theta_{D-3} d\theta_{D-3}
\int^{2\pi}_0 d\theta_{D-2} .
\end{equation}
We can choose the space coordinate 
in the direction 
$\boldsymbol{p}=(p, 0, 0, \cdots)$, 
 then $\boldsymbol{k}\cdot\boldsymbol{p}=kp\cos\theta_1$. The integrand in 
Eq.(\ref{pole:pi:ftfm:3}) depends only on $k$ and $\theta_1$. 
The angle integrals 
with respect to $\theta_2\cdots\theta_{D-2}$ give  
\begin{equation}
\int^{\pi}_0
\sin^{D-4}\theta_2 d\theta_2 \cdots \int^{\pi}_0 
\sin \theta_{D-3} d\theta_{D-3}
\int^{2\pi}_0 d\theta_{D-2} =\frac{2\pi^{D/2-1}}{\Gamma(D/2-1)} .
\end{equation}
Thus the Eq.(\ref{pole:pi:ftfm:4}) reduces to
\begin{eqnarray}
   0 &=& \frac{1}{2g_\pi^0} \left(1-\frac{\langle\sigma\rangle}{\langle\sigma\rangle+m}\right)
\nonumber \\ 
  && +
  \frac{2N_c (p_0^2-\boldsymbol{p}^2)}{(2\pi)^{D/2}\Gamma(D/2-1)} 
          \int_0^\infty k^{D-2} dk \int_0^\pi \sin^{D-3}\theta_1 d\theta_1
\nonumber \\ 
  &&\times
  \left[\left(\frac{1}{E_1}+\frac{1}{E_2}\right)
\right. 
  \frac{1-n_+(E_1)-n_-(E_1)}
       {p_0^2-(E_1+E_2)^2-i\epsilon}
\nonumber \\ &&
  +\left(\frac{1}{E_1}-\frac{1}{E_2}\right)
\left.
  \frac{-n_+(E_1)-n_-(E_1)}
       {p_0^2-(E_1-E_2)^2-i\epsilon}
  \right] .
\label{pole:pi:ftfm:5}
\end{eqnarray}
In the chiral limit $m\rightarrow 0$, the solution of this equation is 
 $p_0^2-\boldsymbol{p}^2=0$, as is shown before. A finite current 
quark mass changes the solution. 

\begin{figure}
\begin{center}
\includegraphics[width=64mm]{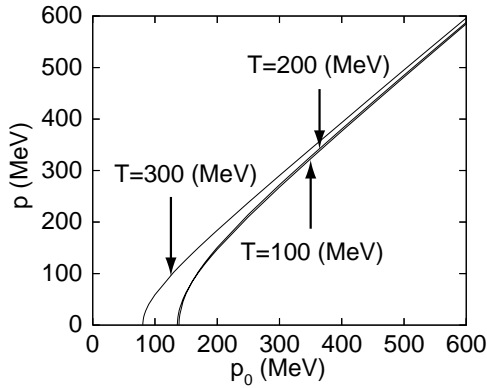}
\end{center}
\caption{Dispersion law for pion at $\mu=0$ for $D=2.4$ and $m_u=m_d=5$MeV.}
\label{dspi}
\end{figure}
\begin{figure}
\begin{minipage}{66mm}
\begin{center}
\vglue 6mm
\includegraphics[width=64mm,height=50mm]{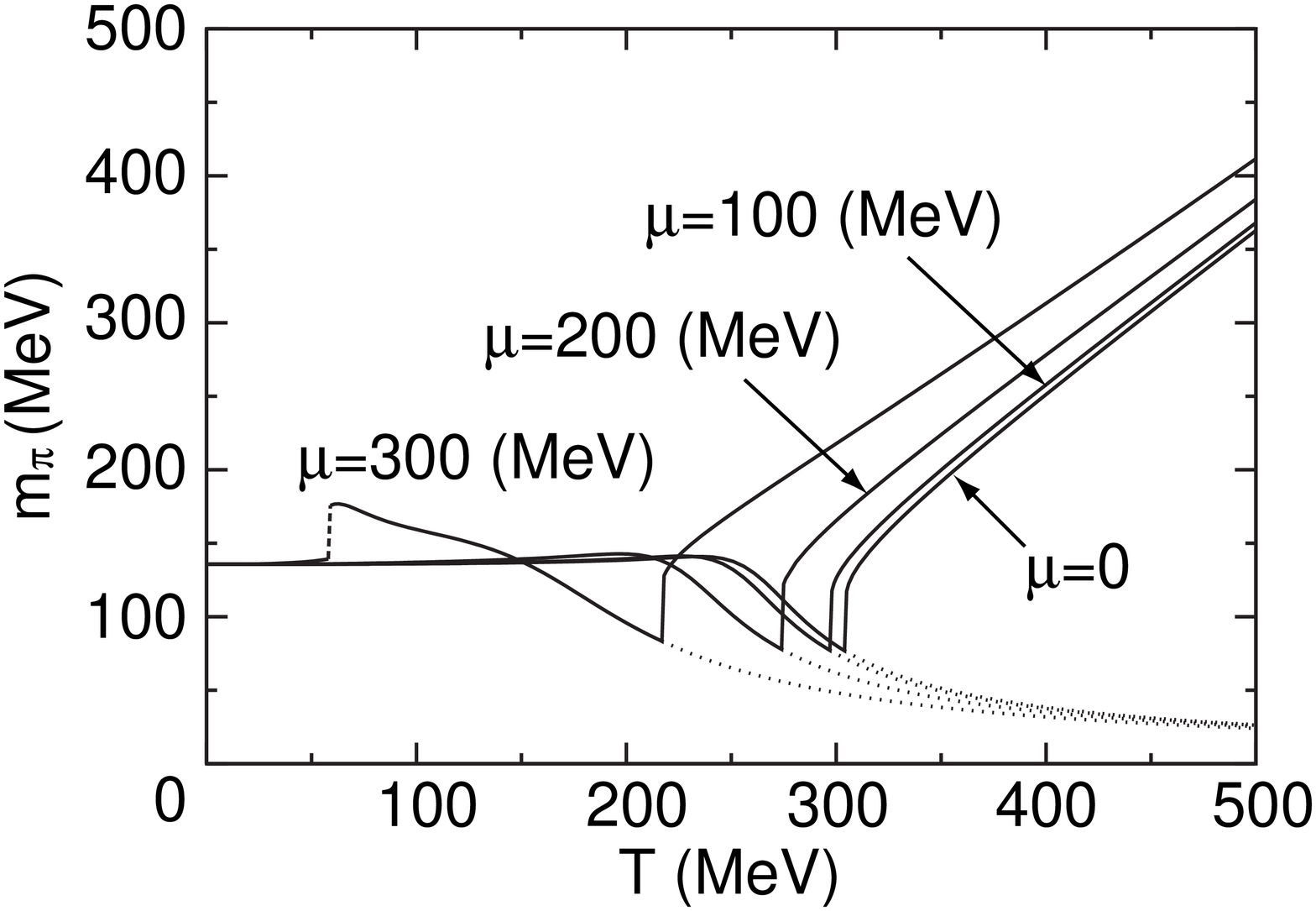}
(a) Full line: Pion mass, dotted line: $-2(\langle \sigma \rangle + m)$.
\end{center}
\end{minipage}
\begin{minipage}{66mm}
\begin{center}
\includegraphics[width=64mm,height=49mm]{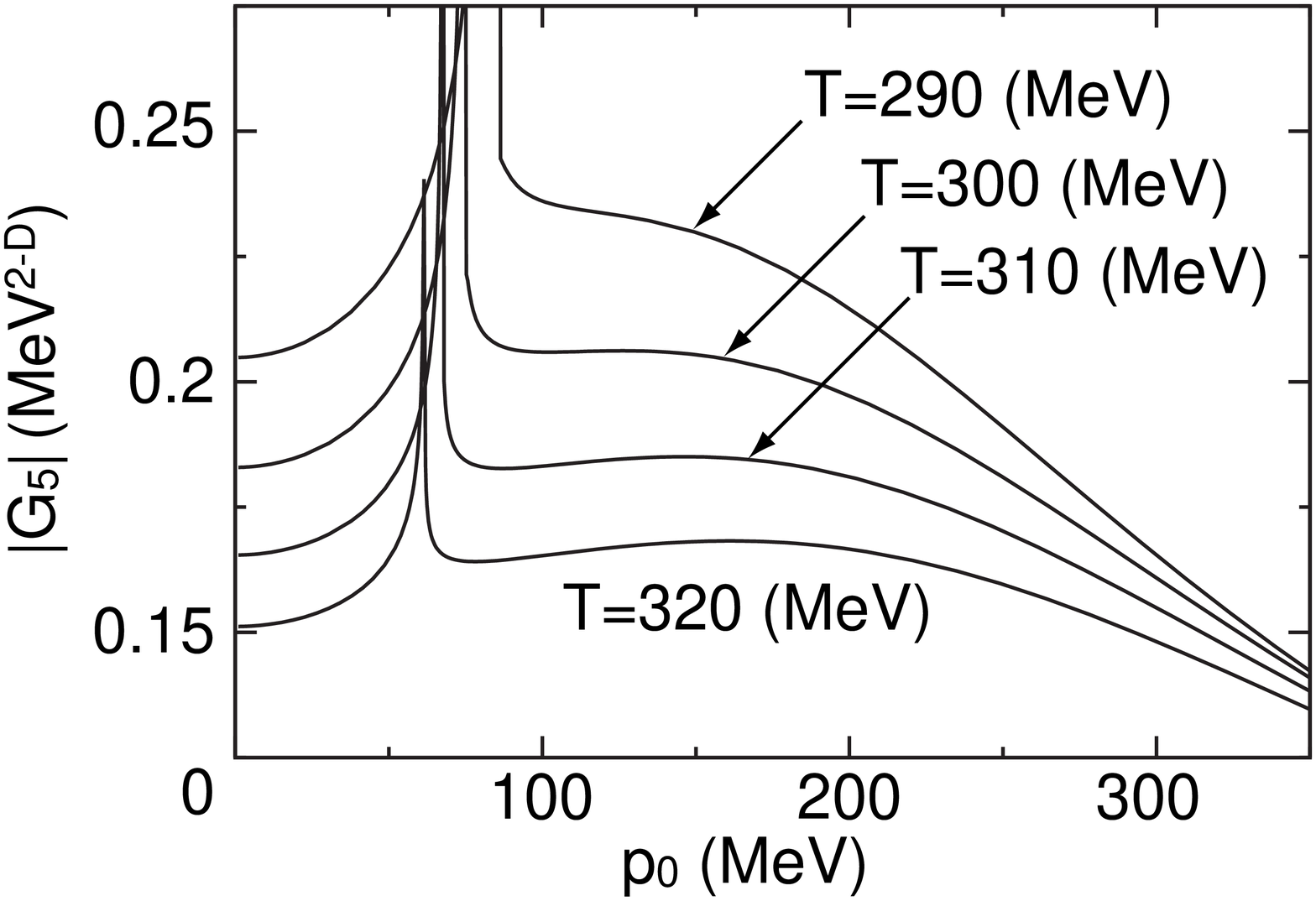}
(b) $|G_5|$ at $\mu=100$MeV.
\end{center}
\end{minipage}
\caption{Typical behavior of the pion pole and $|G_5|$ 
at finite $T$ and $\mu$ for $D=2.4$ and $m_u=m_d=5$MeV.}
\label{m_pi-T}
\end{figure}

In the limit $\mu\rightarrow 0$ and $T\rightarrow 0$, the right hand side 
of Eq.(\ref{pole:pi:ftfm:5}) is a function of $p^2$.
To see if the finite $T$ and $\mu$ modify this dispersion law
we numerically solve Eq.(\ref{pole:pi:ftfm:5}).
In the numerical analysis we put $D=2.4$ and use the parameters determined
in the subsection III.C.
In Fig.~\ref{dspi} we plot the solutions on 
$p_0,\boldsymbol{p}$ plane
at finite $T$. The solutions are on hyperbolic curves as in the
usual $T=0$ case.
The value of $p_0$ on the curve at $\boldsymbol{p}=0$ corresponds to the
pion mass. The pion mass is obtained by solving Eq.~(\ref{pole:pi:ftfm:5})
after rotating the contour by $-\pi / 4$. A soft mode for the pion
channel is found by observing an extremum of the absolute value of the
propagator $|G_5|$.
The pion mass decreases as $T$ increases from 0 
to the critical 
temperature $T_c$. A soft mode appears above $T_c$.
We also draw the behavior of $|G_5|$ near $T_c$ in Fig.~\ref{m_pi-T}
(b). The sharp peak structure is observed near 
$-2(\langle \sigma \rangle + m)$ 
for $T \gtrsim T_c$. In Fig.~\ref{m_pi-T} (a) we see a gap  
at $T \simeq 60$(MeV) for $\mu = 300$(MeV). 
The gap comes from the one for 
$\langle \sigma \rangle$
in Fig.~\ref{fig:potft} (b).
 
For the scalar channel the self-energy, $\Pi_{s}(p_4, \boldsymbol{p})$ 
is given by
\begin{equation}
  \Pi_{s}(p_4, \boldsymbol{p}) = 
  - 4 (g_\pi^0)^2 \frac{1}{\beta} \sum_{n}
  \int \frac{d^{D-1} \boldsymbol{k}}{(2\pi)^{D-1}} \mbox{tr} 
  [S(k)_{\beta\mu}S(k-p)_{\beta\mu}] .
\label{sel:s:ft}
\end{equation}
We perform the trace operation and get
\begin{eqnarray}
\Pi_{s}(p_4, \boldsymbol{p}) &=& 
  4 N_c  (g_\pi^0)^2 2^{D/2} \frac{1}{\beta} 
  \sum_{n} \int \frac{d^{D-1} \boldsymbol{k}}{(2\pi)^{D-1}} \sum_{j\in\{u,d\}}
\label{sel:s:ftp} \\
\nonumber 
&&\times 
  \frac{(\omega_n-i\mu)(\omega_n-p_4-i\mu)+\boldsymbol{k}\cdot(\boldsymbol{k-p})-(m_j+\langle\sigma\rangle)^2}
  {[(\omega_n-i\mu)^2+{E_{1k}}^2][(\omega_n-p_4-i\mu)^2+{E_{2k}}^2]} .
\end{eqnarray}
It is related to the self-energy for the 
pseudo-scalar channel,
\begin{eqnarray}
\Pi_{s}(p_4, \boldsymbol{p})  
&=&\Pi_{5}^3(p_4, \boldsymbol{p})
\nonumber \\
&&   - \sum_{j\in\{u,d\}}8(\langle\sigma\rangle+m_j)^2 N_c  (g_\pi^0)^2 2^{D/2} \frac{1}{\beta} 
  \sum_{n} \int \frac{d^{D-1} \boldsymbol{k}}{(2\pi)^{D-1}}
\nonumber\\ 
&&\times 
  \frac{1}
  {[(\omega_n-i\mu)^2+{E_{1k}}^2][(\omega_n-p_4-i\mu)^2+{E_{2k}}^2]} .
\end{eqnarray}
The difference between $\Pi_{s}(p_4, \boldsymbol{p})$ and 
$\Pi_{5}^3(p_4, \boldsymbol{p})$ disappears at the chiral limit 
$(\langle\sigma\rangle+m_j)\rightarrow 0$.
If the chiral symmetry is restored, the scalar and the pseudo-scalar 
channel degenerate even for finite $T$ and $\mu$.

The two-point function for the scalar meson
can be calculated similarly to Eq.~(63) for the pion pole. After the summation over 
the Matsubara
frequencies and the Wick rotation we obtain
\begin{eqnarray}
   G_s^{-1} (p_0,\boldsymbol{p},\langle \sigma \rangle) 
&=& \frac{1}{2g_\pi^0} \left(1-\frac{\langle\sigma\rangle}{\langle\sigma\rangle+m}\right)
\nonumber\\
  &&   +  \frac{N_c 2^{D/2}(p_0^2-\boldsymbol{p}^2-4(\langle\sigma\rangle+m)^2)}{4} 
          \int \frac{d^{D-1} \boldsymbol{k}}{(2\pi)^{D-1}}
          \frac{1}{E_1E_2}
\nonumber \\ 
  &&\times
  \left[\frac{1-n_+(E_1)-n_-(E_2)}{p_0-E_1-E_2}
       -\frac{n_-(E_1)-n_-(E_2)}{p_0+E_1-E_2}
\right. \nonumber \\ &&\left.  
       +\frac{n_+(E_1)-n_+(E_2)}{p_0-E_1+E_2}
       -\frac{1-n_-(E_1)-n_+(E_2)}{p_0+E_1+E_2}
  \right] .
\label{pole:sig:ftfm}
\end{eqnarray}

Adopting the 
 mathematical trick used in Eq.(\ref{pole:pi:ftfm:4}) and
integrating over the angle variables, $\theta_2\cdots\theta_{D-2}$, we
get
\begin{eqnarray}
 G_s^{-1} (p_0,\boldsymbol{p},\langle \sigma \rangle)
   &=& \frac{1}{2g_\pi^0} \left(1-\frac{\langle\sigma\rangle}{\langle\sigma\rangle+m}\right)
\nonumber \\ 
  &&\hspace{-5mm} +
  \frac{2N_c (p_0^2-\boldsymbol{p}^2-4(\langle\sigma\rangle+m)^2)}{(2\pi)^{D/2}\Gamma(D/2-1)} 
          \int_0^\infty k^{D-2} dk \int_0^\pi \sin^{D-3}\theta_1 d\theta_1
\nonumber \\ 
  &&\times
  \left[\left(\frac{1}{E_1}+\frac{1}{E_2}\right)
\right. 
  \frac{1-n_+(E_1)-n_-(E_1)}
       {p_0^2-(E_1+E_2)^2-i\epsilon}
\nonumber \\ &&
  +\left(\frac{1}{E_1}-\frac{1}{E_2}\right)
 \left.
  \frac{-n_+(E_1)-n_-(E_1)}
       {p_0^2-(E_1-E_2)^2-i\epsilon}
  \right] .
\label{pole:sig:ftfm:5}
\end{eqnarray}

\begin{figure}
\begin{minipage}{66mm}
\begin{center}
\vglue 16mm
\includegraphics[width=64mm,height=50mm]{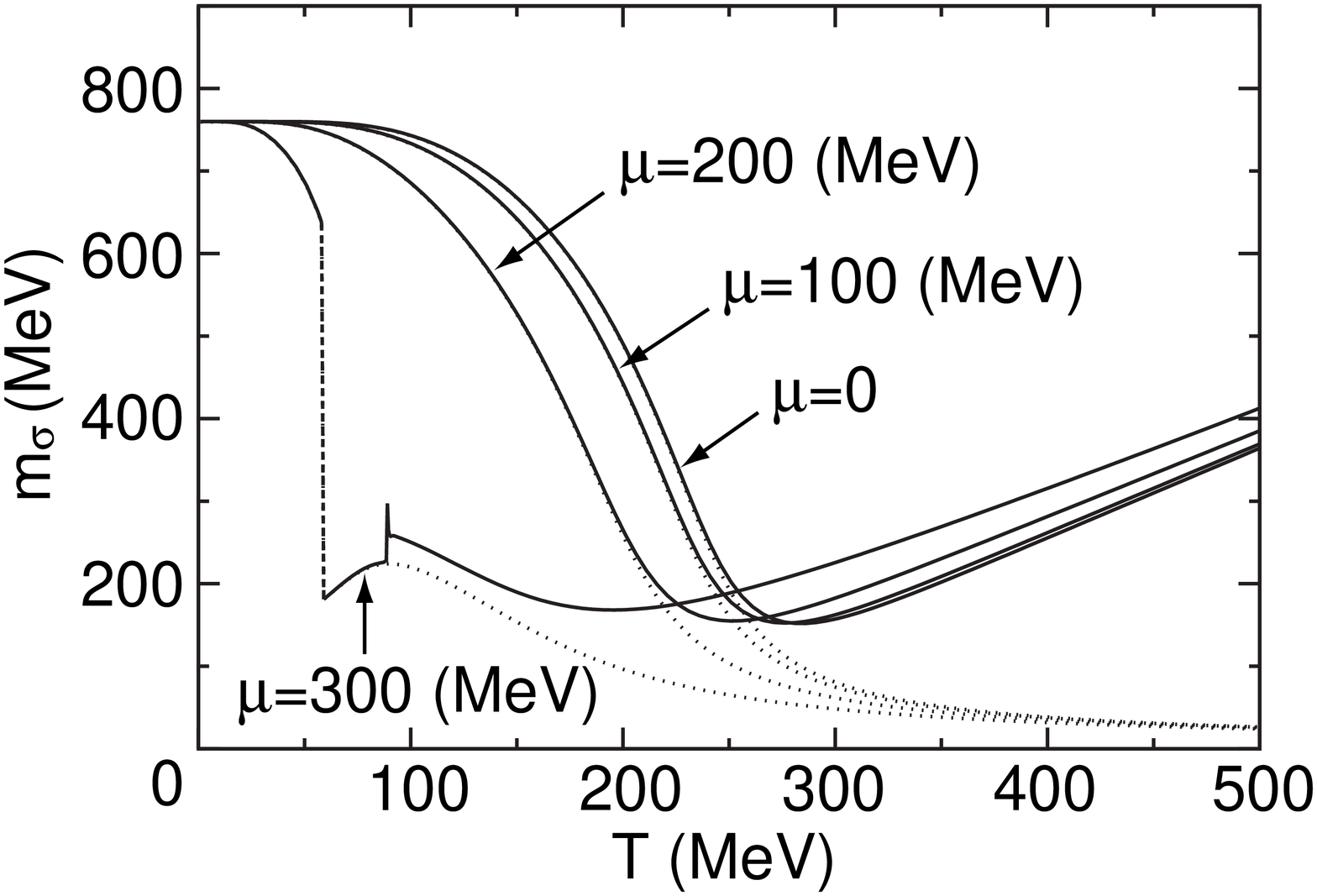}
(a) Full line: sigma meson and soft mode masses, 
dotted line: $-2(\langle \sigma \rangle +m)$.
\end{center}
\end{minipage}
\begin{minipage}{66mm}
\begin{center}
\includegraphics[width=64mm,height=50mm]{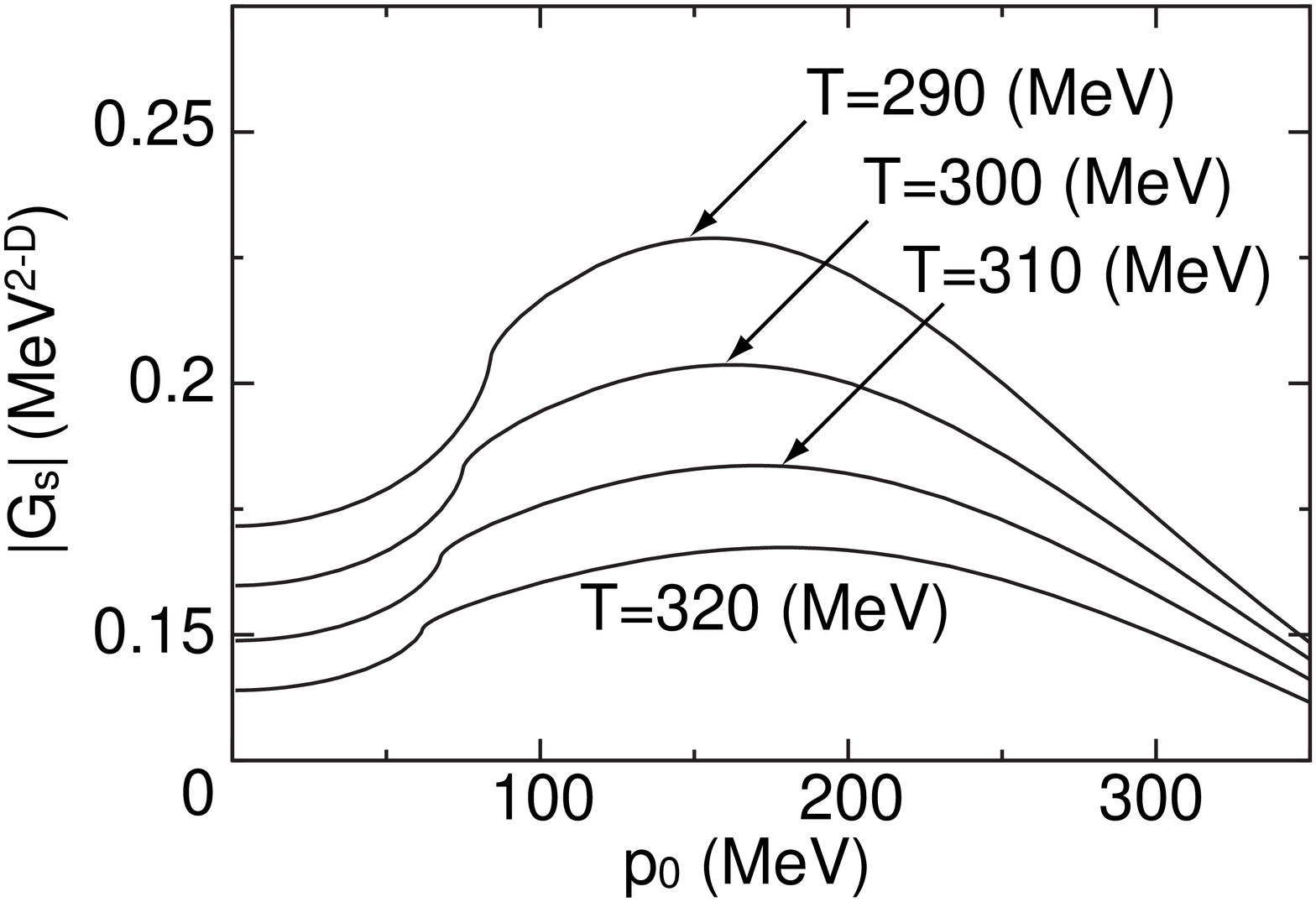}
(b) $|G_s|$ at $\mu=100$MeV.
\end{center}
\end{minipage}
\caption{Typical behavior of the sigma and pion pole at finite $T$ and $\mu$ for $D=2.4$ and $m_u=m_d=5$MeV.}
\label{m_sig-T}
\end{figure}

The sigma meson and soft mode masses are obtained by observing the 
extremum of $|G_s|$
after rotating the contour by $-\pi / 4$.
We draw the behavior of
the sigma meson mass and the soft mode in Fig.~\ref{m_sig-T} (a).
As temperature increases from 0 to $T_c$, the sigma meson mass 
decreases for 
$\mu \lesssim 200$(MeV).
The sigma meson mass 
turns into the soft mode at temperature above
$T_c$.  A gap at $\mu = 300$(MeV) and 
$T \simeq 60$(MeV) in Fig.~\ref{m_sig-T} (a) is induced 
by the gap found in the $<\sigma>$ behavior shown in Fig.~\ref{fig:potft} (b).
The other small gap appears at $T \simeq 100$(MeV). 
It comes from the extremum of 
$<\sigma>$ as is shown in Fig~6. (b). 
In Fig.~\ref{m_sig-T} (a) we also draw  
$-2(\langle \sigma \rangle +m)$ where a naive threshold (at double
constituent quark mass) for $\bar\psi \psi$ state appears.
The sigma channel above $T_c$ we call the soft mode. In Fig.~9 (b)
we show the extrema for $|G_s|$ which corresponds to the soft mode.

The sigma and pion mass has been analysed in \cite{SMMR} at finite $T$
and $\mu$ in the cut-off regularization. For small $\mu$
a similar behavior is observed across the crossover transition. 
For larger $\mu$ the first order phase transition takes place in
both regularizations. In this case the constituent quark mass
$\langle \sigma \rangle$ shows some different behavior. 
For instance, the small peak in the constituent
quark mass appears at $\mu = 300$(MeV) and $T \simeq 100$(MeV) 
in Fig.~\ref{m_sig-T} (a) in the dimensional regularization.

\section{Conclusion}

We studied characteristic features of the NJL model in the dimensional 
regularization. The dimensional regularization is applied to 
momentum integrals for internal fermion lines as usual. Since the model is not
renormalizable, we can not take the four dimensional limit. 
We evaluated some physical properties of the model in the space-time
dimensions less than four. 
We take notice of that only the radiative 
corrections should be evaluated in the space-time dimensions less than four 
to keep the four-dimensional properties in the real world.
The classical parts are evaluated in four dimensions.

In the NJL model we see that the approximate chiral symmetry is dynamically 
broken for a negative bare coupling. The constituent quark mass 
is also negative. 
After the renormalization, we can define the 
renormalized coupling constant.
The dimensional regularization keeps most of 
symmetries. We have inspected
the massless pole of the Nambu-Goldstone mode of the chiral symmetry
breaking at the massless quark limit. Calculating the 
meson masses in
the vacuum, we show that three pseudo-scalar modes have massless pole in the
broken phase. It is the direct consequence of the Ward-Takahashi identity. 
In the dimensional regularization the Ward-Takahashi identity 
remains usual. We also see that the mass for the scalar mode is real and positive. 

The coupling constant, the space-time dimensions and the renormalization
scale can be fixed by the pion mass and 
its decay constant. 
The dimensions 
less than three, but more than two, are preferable to obtain the constituent 
quark mass about $300\sim 400$MeV. The current quark mass has a 
non-negligible dependence on it. 
If we take the larger current quark mass, the constituent quark mass 
about $300$MeV is realized for  higher dimensions. However, 
a large chemical potential induces the first order phase transition 
for lower dimensions $2\leq D\leq 3$ 
in the massless quark limit, as is the case in 
other approaches, such as the cut-off regularization \cite{HK, BMZ, AY}, 
Schwinger-Dyson equation \cite{TY, HS} and the lattice QCD \cite{FK}.

After we fix the parameters phenomenologically, we numerically evaluate 
the constituent quark mass at finite $T$ and $\mu$. The first order 
transition is observed for $D=2.4$ at $\mu=300$MeV. We also evaluate
the thermal influence on the meson masses 
in the leading order of $1/N$ expansion. 
We show the dispersion law for pion and the
behavior of meson 
masses. For lower temperature only a small influence
is observed on the meson mass. It should be noted that boson loops have 
$O(T^2)$ contribution at finite temperature. A meson loop appears 
in the next to leading order of $1/N$ expansion. Though it is neglected 
in the present paper, the contribution from a meson loop may not be 
negligible at higher temperature. 

Some problems remain. 
One should include the QED corrections
to evaluate the neutral and the charged pion mass difference \cite{FIK}. 
The color superconductivity is also interesting to consider in the extended 
NJL model by using the dimensional regularization \cite{FIK2}. \\

\begin{acknowledgments}
The authors would like to thank M. Harada and T. Fujihara 
for useful discussions. T. I. is supported by the Ministry of Education,
Science, Sports and Culture, Grant-in-Aid for Scientific Research (C),
18540276, 2007. A. K. is supported by the GNSF grant No GNSF ST06/4-050.
\end{acknowledgments}

\end{document}